\documentclass[a4paper,11pt]{article}

\usepackage{jheppub} 
\usepackage[normalem]{ulem}  

\usepackage{subcaption}

\usepackage{diagbox}

\usepackage{cleveref}

\newcommand{\Hi}{H_{\rm inf}}
\newcommand{\Pcal}{{\mathcal{P}}}

\newcommand{\lran}[1]{{\langle {#1}\rangle}}

\newcommand{\Mpl}{{M_{\text{pl}}}}

\graphicspath{{./Figures/}}

\title{Cosmological Signatures of Superheavy Dark Matter}

\author[a,b]{Lingfeng Li,}

\author[a,b]{Shiyun Lu,}

\author[a,b]{Yi Wang,} 

\author[c]{and Siyi Zhou}

\affiliation[a]{Department of Physics,
\newline
Hong Kong University of Science and Technology, Hong Kong, P.R.China}

\affiliation[b]{HKUST Jockey Club Institute for Advanced Study,
\newline
 Hong Kong University of Science and Technology, Hong Kong, P.R.China}

\affiliation[c]{The Oskar Klein Centre for Cosmoparticle Physics \& Department of Physics, 
\newline
Stockholm University, AlbaNova, 106 91 Stockholm, Sweden}

\emailAdd{iaslfli@ust.hk}

\emailAdd{sluah@connect.ust.hk}

\emailAdd{phyw@ust.hk}

\emailAdd{siyi.zhou@fysik.su.se}

\abstract{We discuss two possible scenarios, namely the curvaton mechanism and the dark matter density modulation, where non-Gaussianity signals of superheavy dark matter produced by gravity can be enhanced and observed. In both scenarios, superheavy dark matter couples to an additional light field as a mediator. In the case of derivative coupling, the resulting non-Gaussianities induced by the light field can be large, which can provide inflationary evidences for these superheavy dark matter scenarios.}

\begin{document} 
\setcounter{tocdepth}{2}
\maketitle
\flushbottom
 
\section{Introduction}
The nature of dark matter (DM) is a frontier in the modern cosmology. There have been plenty of astronomical and cosmological evidences for DM, such as the galaxy rotation curve, virial velocities of galaxy clusters, gravitational lensing, bullet clusters, supernovae, cosmic microwave background, existence of galaxies in lifetime of the universe and existence of galaxies on scale of milky way. 

The production mechanism of the dark matter is not known so far. It can be produced during inflation or subsequent cosmological evolutions. If the dark matter is produced during inflation, it is hopeful to use inflation as an avenue to probe the nature of the dark matter. There are multiple possible mechanisms. One of them is that gravitational production \cite{Ford:1986sy} during inflation. Relevant models include Planckian Interacting Dark Matter (PIDM) \cite{Garny:2015sjg,Garny:2017kha,Hashiba:2018iff,Haro:2018zdb,Hashiba:2018tbu}, WIMPZILLA \cite{Kolb:1998ki,Kolb:2007vd}, SUPERWIMP \cite{Feng:2010gw}, FIMP \cite{Hall:2009bx} and so on. They can be scalar \cite{Tang:2016vch,Ema:2016hlw}, vector, fermion or spin-2 particles \cite{Babichev:2016bxi} from different types of beyond standard model physics. Such dark matter candidates are usually hard to probe using collider experiments due to the large mass and small coupling (the gravitationally produced dark matter even do not have electroweak interactions with standard model particles). 

It is thus interesting to study the dark matter properties, such as mass, width, spin, parity and couplings from cosmology, using the method of cosmological collider physics \cite{Chen:2009we,Baumann:2011nk,Noumi:2012vr,Arkani-Hamed:2015bza}, which has attracted much attention recently \cite{Chen:2009zp,Assassi:2012zq,Sefusatti:2012ye,Norena:2012yi,Emami:2013lma,Liu:2015tza,Dimastrogiovanni:2015pla,Schmidt:2015xka,Chen:2015lza,Bonga:2015urq,Delacretaz:2015edn,Flauger:2016idt,Lee:2016vti,Delacretaz:2016nhw,Meerburg:2016zdz,Chen:2016uwp,Chen:2016hrz,An:2017hlx,Tong:2017iat,Iyer:2017qzw,An:2017rwo,Kumar:2017ecc,Riquelme:2017bxt,Saito:2018omt,Cabass:2018roz,Dimastrogiovanni:2018uqy,Bordin:2018pca,Arkani-Hamed:2018kmz,Kumar:2018jxz,Goon:2018fyu,Wu:2018lmx,Chua:2018dqh,Wang:2018tbf,McAneny:2019epy,Li:2019ves,Kim:2019wjo,Sleight:2019mgd,Biagetti:2019bnp,Sleight:2019hfp,Welling:2019bib,Alexander:2019vtb,Lu:2019tjj,Hook:2019zxa,Hook:2019vcn,ScheihingHitschfeld:2019tzr,Baumann:2019oyu,Wang:2019gbi,Liu:2019fag,Wang:2019gok,Wang:2020uic}. From the frequency of the oscillation on the squeezed limit non-Gaussianity, one can directly obtain information about the mass of extra fields during inflation. We investigated the search for the dark matter mass in the context of cosmological collider physics in \cite{Li:2019ves}, targeting at the class known as superheavy dark matter (SHDM) \cite{Chung:1998is,Kuzmin:1998uv,Chung:1999ve,Kuzmin:1999zk,Chung:2001cb} with mass $m_{\sigma} \geq H$. Since the dark matter is produced gravitationally, the signal is subject to a suppression of power $(H/M_{\rm pl})$, which makes it hardly observable in the near future experiments.

In this paper, we would like to investigate scenarios with signals within the reach of the near future experiments on primordial non-Gaussianities, such as CMB-S4 \cite{Abazajian:2019eic}, Simons Observatory \cite{Ade:2018sbj}, DESI \cite{Aghamousa:2016zmz}, EUCLID \cite{Amendola:2016saw}, SPHEREx \cite{Dore:2014cca} and LSST \cite{Abell:2009aa}. The idea utilizes the curvaton scenario \cite{Enqvist:2001zp, Lyth:2001nq, Moroi:2001ct}, where more than one light field is present during inflation. The curvatons are subdominant in the energy density during inflation. In the subsequent evolutions, they are the main source of the primordial curvature perturbations. The dark matter field, on the other hand, is another type of field which has mass $m_{\sigma}  \geq H$. As we will see, this scenario can give larger non-Gaussianities (NG's) on the primordial curvature perturbations. Such possibility of large NG from the curvaton scenario has also been studied in \cite{Kumar:2019ebj,Wang:2019gbi}. Another possibility to imprint the dark matter information through a light field is modulated production of dark matter, where the dark matter mass, and thus the production rate, is modulated by a light scalar. This is related to the mechanism of modulated reheating~\cite{Dvali:2003em, Kofman:2003nx}.

This paper is organized as follows: in Sec.~\ref{model}, we setup the model which consists of an inflaton, a massive scalar field and a light field. In Sec.~\ref{DMproduction}, we discuss the production of dark matter during inflation. In Sec.~\ref{iso}, we calculate the primordial three-point and four-point correlation functions of the light field. In Sec.~\ref{curvaton} and Sec.~\ref{spectator}, we discuss the post-inflationary evolutions of the light field and their observational constraints. More specifically, Sec.~\ref{curvaton} devotes to the case where the power spectrum is mostly generated by $\chi$. Sec.~\ref{spectator} devotes to the case where the power spectrum is mostly generated by the inflaton field. We give a brief summary in the end in Sec.~\ref{summary}. 

\section{Model}\label{model}
We consider a cosmology model with an inflaton $\phi$, a heavy field $\sigma$ with mass $m_{\sigma}\gtrsim \Hi$, and a light field $\chi$ with mass $m_\chi \ll m_\phi$ with an action
\begin{align}\nonumber
	S = -\int d^{4} x \sqrt{-g}\bigg[ & \frac{1}{2}\left(\partial_{\mu} \phi\right)^{2}+\frac{1}{2} m_{\phi}^{2} \phi^{2}+\frac{1}{2}\left(\partial_{\mu} \sigma\right)^{2}+\frac{1}{2} m_{\sigma}^{2} \sigma^{2}+\frac{1}{2}\left(\partial_{\mu} \chi\right)^{2}    +V(\chi) \bigg]+S_{\rm int}~.
\end{align}
Both the inflaton field $\phi$ and the curvaton $\chi$ are light and have rolling background. The heavy field $\sigma$ does not have a VEV. Also, the inflaton does not couple to either the heavy field $\sigma$ or the light field $\chi$ while all fields couples to gravity minimally for simplicity. We can decompose the light field $\chi$ into a time-dependent background and perturbations
as $\chi (t,\mathbf x) = \bar \chi(t) + \delta \chi (t, \mathbf x)$. We further assume that the background of light field varies slowly during inflation, such that it approximates to a constant during inflation $\bar \chi (t) \simeq \bar\chi$. Aside from these requirements, the form of $\chi$ potential $V(\chi)$ is left free as long as $\rho_\chi$ becomes insignificant compared to other components in the late time universe. 

We discuss two possibilities for the interaction term $S_\mathrm{int}$, namely direct coupling and derivative coupling. In the case of direct coupling,
\begin{align}
	S_{\rm int} = -\int d^{4} x \sqrt{-g}\bigg[ \frac{1}{4} \lambda \sigma^2 \bar\chi^2 + \frac{1}{2} \lambda \bar \chi \sigma^2  \delta \chi + \frac{1}{4} \lambda \sigma^2 \delta \chi^2  \bigg]~.
\label{eq:model}
\end{align}
the first term and the third term contribute to the effective mass of the heavy field $\sigma$, the second term and third term are interaction terms between the heavy field and $\chi$. 
\begin{align}
	m_{\sigma \rm eff}^2 = m_{\sigma}^2 + \frac{1}{2} \lambda \bar\chi^2  + \frac{1}{2} \lambda \langle \delta \chi^2 \rangle~. 
	\label{eq:meff}
\end{align}
On the other hand, it is also well motivated to adopt the derivative $\sigma-\chi$ coupling of the form $\frac{1}{4}\lambda_2 \sigma^2 \dot\chi^2$, leading to interaction 
\begin{align} 
	S_{\rm int}^\prime = -\int d^{4} x \sqrt{-g}\bigg[ \frac{1}{4} \lambda_2 \sigma^2 \dot{\bar\chi}^2 + \frac{1}{2} \lambda_2 \dot{\bar \chi} \sigma^2  \delta \dot\chi + \frac{1}{4} \lambda_2 \sigma^2 (\delta \dot\chi)^2  \bigg]~.
\label{eq:model-prime}
\end{align}
the first term and the third term contribute to the effective mass of the heavy field $\sigma$, the second term and third term are interaction terms between the heavy field and $\chi$. 
\begin{align} 
	m_{\sigma \rm eff}^2 = m_{\sigma}^2 + \frac{1}{2} \lambda_2 \dot{\bar\chi}^2  + \frac{1}{2} \lambda_2 \langle (\delta \dot\chi)^2 \rangle~. 
\end{align}

For the $\chi$ field, the background $\bar \chi$ is governed by the following equation of motion
\begin{align}
\label{eq:chislowroll}
	\ddot {\bar \chi} + 3 H \dot {\bar \chi }+ m_{\chi}^2 \bar \chi = 0~,
\end{align} 
which holds during the post inflation era as well. Admitting the $\chi$ equation of motion Eq.~\ref{eq:chislowroll} and imposing $\ddot{\chi}\sim 0$, we can get the $\chi$ change rate during the primordial era as:
\begin{equation}
\dot {\bar \chi } = -\frac{m_\chi^2}{3 H} \bar{\chi}~.
\end{equation}

The inflaton is the field that drives inflation. In the rest of this paper, we assume that the change of energy density contributed from  $\chi$ subdominates during inflation, i.e. $\chi$ is a spectator field during inflation. In other words, the following constraint is satisfied
\begin{align}
	\dot \rho_{\chi\rm inf} \ll \dot \rho_{\phi\rm inf}~, 
\end{align}
where $\rho_{\chi\rm inf}$ and $\rho_{\phi\rm inf}$ denote the energy density of the light field $\chi$ and the inflaton $\phi$ during inflation, respectively.

The observable of interest is the primordial three-point correlation or four-point correlation of the primordial curvature perturbation $\zeta$, defined by the following metric written in the $\zeta$ gauge where there is no energy density fluctuation,
\begin{align}
	ds^2 = -dt^2 + a^2 (t) (1+2\zeta) d x^i dx^j ~. 
\end{align}

\section{Superheavy DM Gravitational Production During Inflation}
\label{DMproduction}
From our model~\eqref{eq:model} it is obvious that the heavy field $\sigma$ is protected by the $Z_2$ symmetry and only interacts with the $\chi$ field. As discussed before, without further interactions $\sigma$ is then a massive stable particle and interacting with other fields via the small coupling and $\chi$ as the mediator, making itself a good DM candidate. For its gravitational production during inflation, we will mostly follow the calculations in~\cite{Li:2019ves}.

During inflation and in the presence of a background $\chi$ field $\sim\bar{\chi}$, the equation of motion for $\sigma$ is
\begin{align}
	\ddot \sigma + 3 H \dot \sigma - \frac{1}{a^2} \nabla^2 \sigma + m_{\sigma{\rm eff}}^2 \sigma =0~,
\end{align}
where $m_{\sigma{\rm eff}}$ is given by ~\eqref{eq:meff}. During the inflationary era, the field $\sigma$ can be expressed in the standard formalism (see also \cite{Markkanen:2016aes}):
\begin{align}
\sigma = \int \frac{d^3 \mathbf k}{(2\pi)^3} e^{i \mathbf k\cdot \mathbf x} a^{-3/2} [f_k a_{\mathbf k} + f_k^* a_{-\mathbf k}^\dagger ]~,
\end{align}
where $a_\mathbf k$ and $a_{-\mathbf k}^\dagger$ are the annihilation and creation operators that satisfy the commutation relations $[a_{\mathbf k}, a_{\mathbf k'}] = 0$ and $[a_{\mathbf k}, a_{\mathbf k'}^\dagger] = (2\pi)^3 \delta^{(3)} (\mathbf k - \mathbf k')$. We then find
\begin{align}
	 \ddot f_k(t)  + \omega_k^2 f_k (t) = 0, \quad \omega_k^2=  \frac{k^2}{a^2} + H^2 \mu^2 - \frac{3}{2} \dot H , \quad \mu\equiv \sqrt{\frac{m_{\sigma{\rm eff}}^2}{H^2}-\frac{9}{4}} ~.
\end{align}
Since we assume minimal coupling of $\sigma$, the $\xi R$ term is zero in the definition of $\mu$. During slow roll inflation, the scale factor is approximately $a(t) = e^{\Hi t} = -1/(\Hi \tau)$, and $R = 12 \Hi^2$ and $\dot \Hi$ vanishes. Hence we refer $\Hi$ as $H$ in the following discussions. When inflation ends, the scale factor starts to evolve in a non-accelerated way. There are two classes of solutions to the massive field equation of motion, corresponding to ``in" state and ``out" state, respectively \cite{Chen:2009we}:
\begin{align}
	f_k^{\rm in} (t)  = \sqrt{\frac{\pi}{4H}} e^{-\pi\mu/2} H_{i\mu}^{(1)} (-k\tau)~, 
	f_k^{\rm out} (t)  = \bigg(\frac{2H}{k}\bigg)^{i\mu} \frac{\Gamma(1+i\mu)}{\sqrt{2H\mu}} J_{i\mu} (-k\tau)~, \label{outstatemodefunction}
\end{align}
where $H^{(1)}_{\nu}(x)$ is the Hankel function of the first kind, and $J_{\nu}(x)$ is the Bessel function. 

The mode functions are related via a Bogoliubov transformation as
\begin{align}\label{eq:expand_Bogoliubov}
	f_k^{\rm in} (t) = \alpha_k f_k^{\rm out} (t) + \beta_k f_k^{\rm out *} (t) ~.
\end{align}
Inserting the explicit expressions for the ``in" state mode function and ``out" state mode function, we obtain the following expressions for the Bogoliubov coefficients
\begin{align}
	\beta_k = \bigg(\frac{2H}{k}\bigg)^{i\mu} \frac{e^{\pi\mu/2}\sqrt{2\pi\mu}}{(1-e^{2\pi\mu})\Gamma[1-i\mu]}~, \quad \alpha_k = - e^{\pi\mu} \beta_k^*~,  
\end{align}
with the outgoing amplitude $|\beta_k|^2=1/(e^{2\pi\mu}-1)$. The comoving number density of $\sigma$ DM produced in the pure slow roll approximation then reads 
\begin{align}\label{num_density1}
	N_\sigma = \int_{0}^{\infty}  dk \,\, 2\pi  k^2 |\beta_k|^2~,
\end{align}
which gives the total number of particles from the past infinity to future infinity, as shown below (see also \cite{Kobayashi:2014zza}). Since particles are produced when $|\omega_k'/\omega_k^2|$ takes its maximum and thus $\tau\sim -\mu/k$ for each mode. The $k$ integral is then rewritten as:
\begin{align}
	\int_0^\infty dk\,\, k^2 = \mu^3 \int_{-\infty}^0 d\tau\,\, \bigg(-\frac{1}{\tau}\bigg)^4 = \mu^3 \int_{-\infty}^0 d\tau\,\,  (aH)^4~.
\end{align}
The corresponding physical number density of particles at each moment can be evaluated as:
\begin{align}
	n_\sigma  =  \frac{1}{a(\tau)^3}2\pi |\beta_k|^2 \mu^3  \int_{-\infty}^{\tau} d\tilde \tau (aH)^4 = \frac{2\pi H^3\mu^3}{3(e^{2\pi\mu}-1)} ~.
\end{align}
which is constant in time. Assuming matter domination before reheating and radiation domination afterwards, the current DM abundance can be expressed numerically~\cite{Chung:2001cb,Li:2019ves}:
\begin{equation}
\Omega_\sigma h^2 \simeq \frac{8\pi}{3}\Omega_{\text{R}}h^2 \frac{m_\sigma n_\sigma}{M_{\rm pl}^2 H^2}\bigg( \frac{T_{\rm RH}}{T_0} \bigg)\simeq 1.14\times 10^{9} \mu^3 e^{-2\pi\mu} \frac{H m_\sigma
 T_{\rm RH}}{M_{\rm pl}^2}
\label{eq:DMdensity1}
\end{equation}
in the unit of GeV. Here $\Omega_{\rm R}$ and $T_0$ are the radiation energy fraction and temperature today. To produce non-negligible DM comparable to the observed value $\Omega_{\rm CDM}h^2 \simeq 0.12$, from the form of \eqref{eq:DMdensity1} it is clear that the model disfavors large $\mu$ and low $H$ or $T_{\rm RH}$. 

In the above slow roll approximation the inflation actually never ends, however, the DM density's exponential dependence on $\mu$ and thus on $\lambda \chi^2$ would be crucial for generating NG in the DM sector. To better estimate the $\sigma$ relic density, we consider the case where inflation is  smoothly connected to Minkowski spacetime. The particle production in this scenario would approximate the particle production in more realistic scenarios where inflation is connected to a stage of the universe with much lower Hubble scale (and thus approximately Minkowski), for example, radiation-dominated universe with transition time of order $H^{-1}$. By introducing the Stokes line method to compute the particle production beyond exponential precision, one obtains the relic density of $\sigma$ today as
\begin{align}
\Omega_\sigma h^2 & \simeq 6.1\times 10^{5}\mu e^{-2\pi\mu} \frac{H  m_\sigma T_{\rm RH}}{M_{\rm pl}^2}~,
\label{eq:OmegaDMStokes}
\end{align}
which is also in the unit of~GeV. The result is similar to ~\eqref{eq:DMdensity1} but with different numerical prefactors and different powers of $\mu$. For moderate $\mu$, the dependence of $n_\sigma$ shall be dominated by the exponent factor. There are other ways to create $\sigma$ through gravity, such as a sudden transition from the inflation phase to the radiation domination phase, which may be the case for models such as brane inflation \cite{Dvali:1998pa,Burgess:2001fx,Dvali:2001fw,Shandera:2003gx,HenryTye:2006uv} or quintessential inflation \cite{Peebles:1998qn}. The $\sigma$ number density in this case can be much larger than the smooth transition case. However, the produced DM number density is proportional to $H^3$ and hence shows little correlation with $\bar{\chi}$. Without the $\mu$ and $\bar{\chi}$ dependence, $\rho_\sigma$ fluctuation cannot carry information of $\chi$ field. Even though, the sudden transition mechanism might be useful in the curvaton case, which we will discuss in Sec.~\ref{curvaton}.

Since the precise $\Omega_\sigma h^2$ is dependent on models of reheating and requires numerical evaluations much more complicated than the semi-analytical approaches we adopt here, it is convenient to parameterize $\Omega_\sigma h^2$ for later use:
\begin{equation}
\Omega_\sigma h^2 = A \mu^\alpha e^{-2\pi \mu} \frac{H  m_\sigma T_{\rm RH}}{M_{\rm pl}^2}~,
\label{eq:DMdensity3}
\end{equation}
where $A$ is an constant between $\mathcal{O}(10^{4-9})$ characterizing the DM production efficiency of different models and $\alpha$ describes the remnant power dependence on $\mu$.

In the above discussions we assume that $\sigma$ particle number is fixed once it is produced near the end of inflation era, one may concern if $\sigma$ annihilates or produced during reheating and radiation domination.  This is indeed a good approximation as long as $\lambda$ is small and $\chi$ only interacts with other fields weakly. Starting from possible $\sigma$ annihilation. For a weakly interacting $\chi$, the process $2\sigma\to 2\chi$ dominates the annihilation, with $s$-wave thermal averaged $\left\langle \sigma v \right\rangle \sim \lambda^2/m_\sigma^2$. For heavy DM like $\sigma$, when the radiation temperature $T\gtrsim m_\sigma$, the physical number density $n_\sigma \ll n_\sigma^{\rm eq} \simeq T^3$, hence annihilation cannot reduce $n_\sigma$. The potential constraint on $\lambda$ and other parameters will be discussed below. When $T\ll m_\sigma$ and thus $n_\sigma^{\rm eq} \simeq T^3 e^{-m_\sigma/T} \ll n_\sigma$, the annihilation can happen, but already freeze out $n_\sigma \left\langle \sigma v \right\rangle \ll H$. The minuscule annihilation rate also ensures $\sigma$ DM is safe against indirect detection bounds, as current indirect detection constraints based on DM annihilation are unable to put constraints on weakly interacting DM ($\left\langle \sigma v \right\rangle \lesssim 10^{-26}$~cm$^3$/s) with mass heavier than $\mathcal{O}(10)$~GeV~\cite{Slatyer:2015jla}.

On the contrary, during the reheating or early stage of radiation domination, the thermal bath can create more $\sigma$ via the mediation of $\chi$. As a result, the constraint on thermal interaction put an upper limit for $\lambda$ if we want to ensure that most $\sigma$ are created by gravity and henceforth $\delta\rho_\sigma$ can carry information of $\chi$ fluctuation rather than that of radiation. In the minimal case, radiation are SM like and massless, created by inflaton decays and interact with $\chi$, allowing the later to decay eventually. The $\chi-$radiation interaction coupling (assuming massless fermions) is then $\sim\mathcal{O}(\sqrt{\Gamma_\chi/m_\chi})\ll 1$. The mass radiation then couples to $\sigma$ in the presence of $\bar{\chi}$ via the effective coupling:
\begin{equation}
\lambda_{\sigma r}\sim\lambda \frac{\sqrt{16\pi m_\chi \Gamma_\chi} \bar{\chi}}{ T^2- m_\chi^2}~.
\label{eq:backreaction}
\end{equation}
As long as $\lambda_ {\sigma r} \ll 1$, the thermal production of $\sigma$ would be suppressed, which is usually satisfied according to our numerical results. Moreover, there are other scenarios where the thermal production of $\sigma$ is further suppressed\footnote{One example is the case that $\chi$ and radiation is not directly coupled. In this scenario, $\chi$ firstly decay to a sector weakly coupled to radiation and the later further decays to radiation. Consequently the radiation created during reheating is not interacting with $\chi$ at leading order.}. For simplicity, here we assume that the constraint of $\lambda_ {\sigma r} \ll 1$ is always satisfied and gravity dominates $\sigma$ production.

\section{Isocurvature Fluctuations During Inflation}\label{iso}
In this section, we consider the NG of $\delta\chi$ during inflation. We consider bispectrum in Section~\ref{bispectrum} and trispectrum in Section~\ref{trispectrum}. We mainly focus on the clock signals which we can use to probe the mass of the superheavy dark matter. 

\subsection{Bispectrum}\label{bispectrum}
\begin{figure}[h!] 
	\centering 
	\includegraphics[width=5cm]{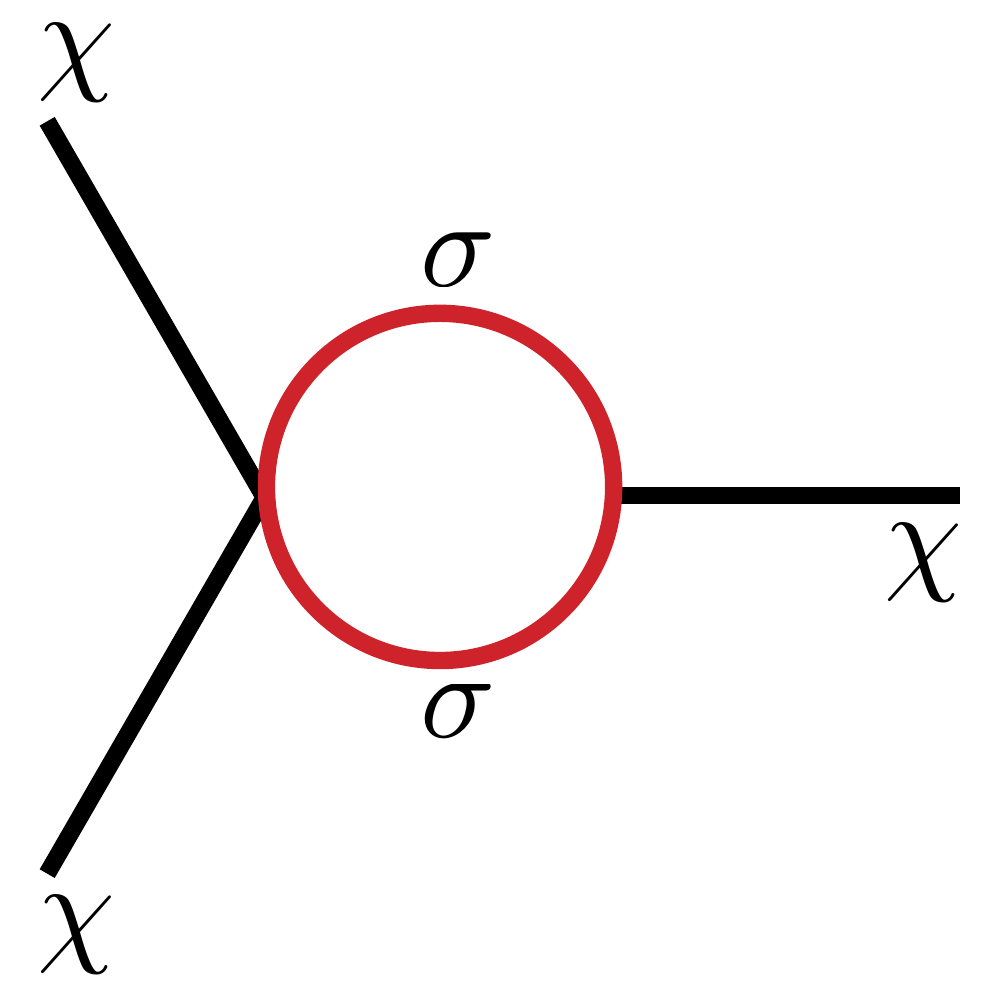} \quad  
	\includegraphics[width=5cm]{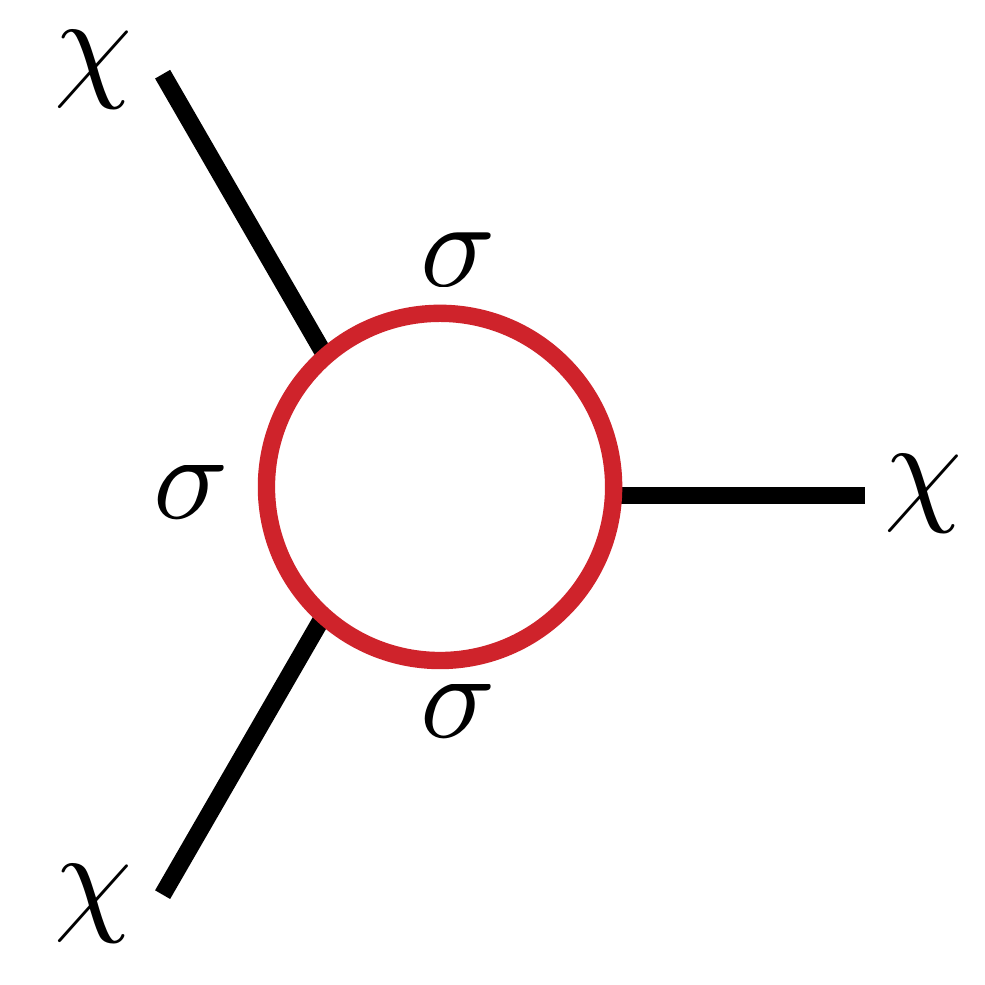}  
	\caption{The diagrams that contributes to $\langle \delta\chi \delta \chi  \delta \chi \rangle$.} \label{bispectrumchichichi}
\end{figure}

The second order action of the primordial curvature perturbation can be written down following the procedure in~\cite{Chen:2010xka,Wang:2013eqj}
\begin{align}
	S_\zeta = M_p^2 \int d t \frac{d^3 k}{(2\pi)^3} \epsilon (a^3 \dot\zeta^2 - k^2 a \zeta^2)~.
\end{align}
Quantizing it in the following way
\begin{align}
	\zeta_{\mathbf k} & = u_k c_{\mathbf k} + u_k^* c^\dagger_{-\mathbf k}  ~,
\end{align}
where $c^\dagger_{\mathbf k}$,  $c_{\mathbf k}$ are the creation and annihilation operators satisfying
the usual commutation relations $  [c_{\mathbf k}, c^\dagger_{\mathbf p}] = (2\pi)^3 \delta^{(3)} (\mathbf k - \mathbf p) $. The mode function satisfies the following equation of motion
\begin{align}
	\ddot \zeta +(3+\eta) H \dot \zeta +\frac{k^2}{a^2} \zeta = 0~.
\end{align}
To the lowest order in slow roll parameter, the solution is
\begin{align}
	u_k (\tau) = \frac{H}{2\sqrt{\epsilon} M_{\rm pl}} \frac{1}{k^{3/2} } (1+i k \tau) e^{-i k \tau}~.
\end{align}
Using the Schwinger-Keldysh formalism, the Feynman diagram on the left hand side of Figure~\ref{bispectrumchichichi} is evaluated as
\begin{align} \nonumber
	& \langle \delta\chi_{\mathbf k_1} \delta\chi_{\mathbf k_2} \delta\chi_{\mathbf k_3} \rangle'_{(2)} = \frac{1}{2} (i \lambda)^2 \bar\chi \sum_{a,b=\pm} ab \int \frac{d \tau_1}{(-H\tau_1)^4}   \frac{d \tau_2}{(-H\tau_2)^4} G_a (k_1;\tau_2) G_a (k_2;\tau_2) G_b (k_3;\tau_1) \\ \label{threepointcorrelationofchi}
	& \times \int \frac{d^3 \mathbf p}{(2\pi)^3}  \int \frac{d^3 \mathbf q}{(2\pi)^3} (2\pi)^3 \delta^{(3)} (\mathbf p+\mathbf q- \mathbf k_3) D_{ab} (p,\tau_1,\tau_2) D_{ab} (q,\tau_1,\tau_2) ~,
\end{align}
where the subscript $(2)$ denotes the number of interaction vertices. This type of integral in the squeezed limit $k_1\simeq k_2\gg k_3$ is well-known~\cite{Arkani-Hamed:2015bza,Chen:2016nrs,Chua:2018dqh,Li:2019ves,Lu:2019tjj} and the details are collected in Appendix.~\ref{loopdetails}. Eq.~\eqref{threepointcorrelationofchi} is then evaluated as
\begin{align}
	& \langle \delta\chi_{\mathbf k_1}\delta \chi_{\mathbf k_2} \delta\chi_{\mathbf k_3} \rangle'_{(2)} =    -\frac{1}{2} \lambda^2 \bar\chi   \frac{1}{  \pi^4  }  \frac{H^2}{64 k_1^3  k_1^3  }   \bigg[ C_{\text{bi}}(\mu) \bigg(\frac{k_3  }{2 k_1}\bigg)^{-2i\mu}  +\text{c.c.} \bigg] ~,
\end{align}
where
\begin{align}
	C_{\text{bi}}(\mu)= \mu^{-2} (2-i\mu)  \Gamma^2 (2-2 i \mu ) \Gamma^2(i\mu) \Gamma^2(\frac{3}{2}-i\mu) \Gamma(4i\mu-4)  \sin(2i\mu\pi)\sin^2(i\mu\pi) ~.
\label{cbispectrum}
\end{align}
The right hand side of Figure~\ref{bispectrumchichichi} is only differed from the Feynman diagram on the left by a factor of $2 \lambda \bar\chi^2/H^2$. So the total contribution to the bispectrum is
\begin{align}
	& \langle \delta\chi_{\mathbf k_1} \delta\chi_{\mathbf k_2} \delta\chi_{\mathbf k_3} \rangle' =     \bigg(1+\frac{2\lambda \bar \chi^2}{H^2}\bigg)   \frac{-\lambda^2 \bar\chi H^2}{128 \pi^4 k_1^6 }   \bigg[ C_{\text{bi}}(\mu) \bigg(\frac{k_3  }{2 k_1}\bigg)^{-2i\mu}  +\text{c.c.} \bigg] ~.
\end{align}

We can replace the interaction vertices in Figure~\ref{bispectrumchichichi} to derivative couplings, resulting in a similar integration:
\begin{align} \nonumber 
	& \langle \delta\chi_{\mathbf k_1} \delta\chi_{\mathbf k_2} \delta\chi_{\mathbf k_3} \rangle'_{(2)} =  \frac{1}{2} (i \lambda_2)^2 \dot{\bar\chi} \sum_{a,b=\pm} ab \int \frac{d \tau_1}{(-H\tau_1)^3}   \frac{d \tau_2}{(-H\tau_2)^2} G^\prime_a (k_1;\tau_2) G^\prime_a (k_2;\tau_2) G^\prime_b (k_3;\tau_1) \\ \label{threepointcorrelationofchid}
	& \times \int \frac{d^3 \mathbf p}{(2\pi)^3}  \int \frac{d^3 \mathbf q}{(2\pi)^3} (2\pi)^3 \delta^{(3)} (\mathbf p+\mathbf q- \mathbf k_3) D_{ab} (p,\tau_1,\tau_2) D_{ab} (q,\tau_1,\tau_2) ~,
\end{align}
where the subscript $(2)$ denotes the number of interaction vertices. Similar to the calculations in the direct coupling case, in the squeezed limit $k_1\simeq k_2\gg k_3$, \eqref{threepointcorrelationofchid} and combine it with the integration of another diagram as:
\begin{align}
	  \langle \delta\chi_{\mathbf k_1} \delta\chi_{\mathbf k_2} \delta\chi_{\mathbf k_3} \rangle' =      \bigg(1+\frac{2\lambda_2 \dot{\bar \chi}^2}{H^2}\bigg)    \frac{-\lambda_2^2 \dot{\bar\chi}H^5}{256  \pi^4  k_1^6}   \bigg[ C_{\text{bi}}^d(\mu) \bigg(\frac{k_3  }{2 k_1}\bigg)^{-2i\mu}  +\text{c.c.} \bigg] ~.
\end{align}
where the loop suppression factor for derivative coupling
\begin{align}
	C_{\text{bi}}^d(\mu)=\Gamma(2-2 i \mu )\Gamma(4-2 i \mu ) \Gamma^2(i\mu) \Gamma^2(\frac{3}{2}-i\mu)\Gamma(4i\mu-4)  \sin(2i\mu\pi)\sin^2(i\mu\pi) ~.
\label{cbispectrumd}
\end{align}

\subsection{Trispectrum}\label{trispectrum}
\begin{figure}[h!] 
	\centering 
	\includegraphics[width=4.5cm]{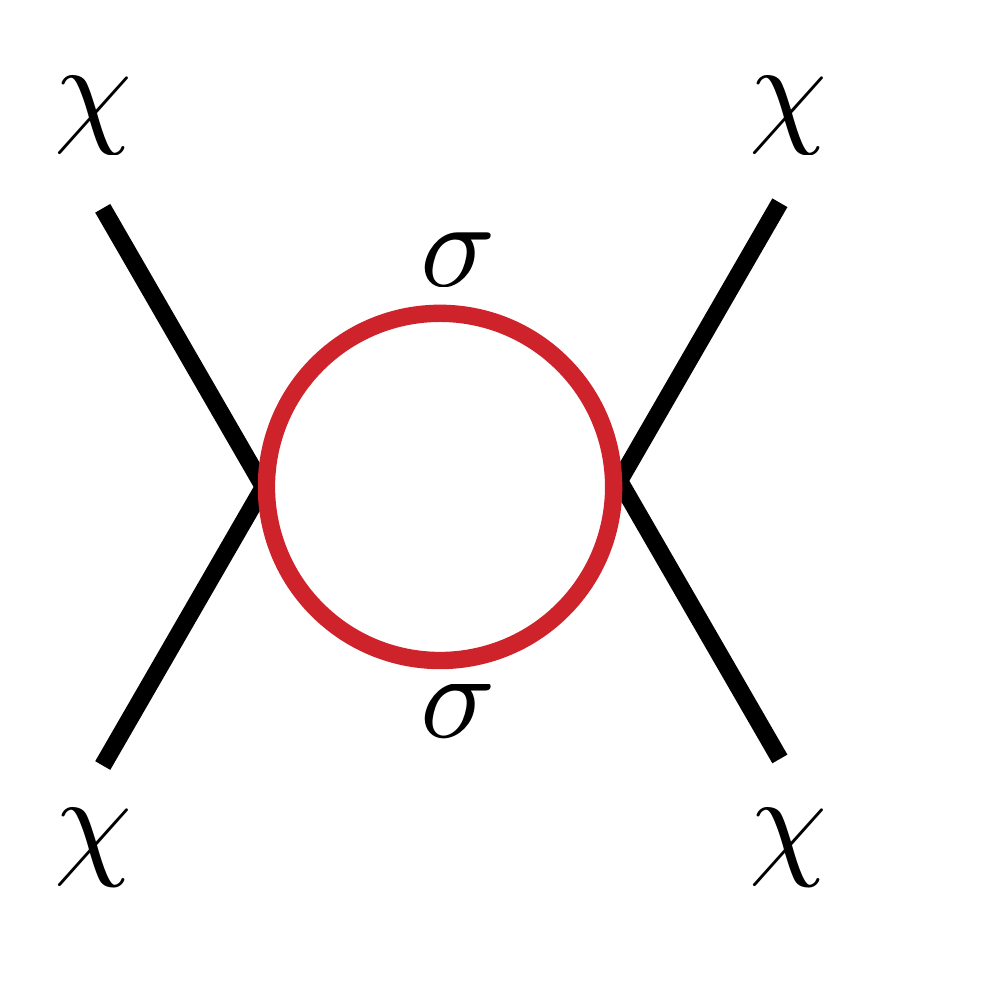}  \quad 
	\includegraphics[width=4.5cm]{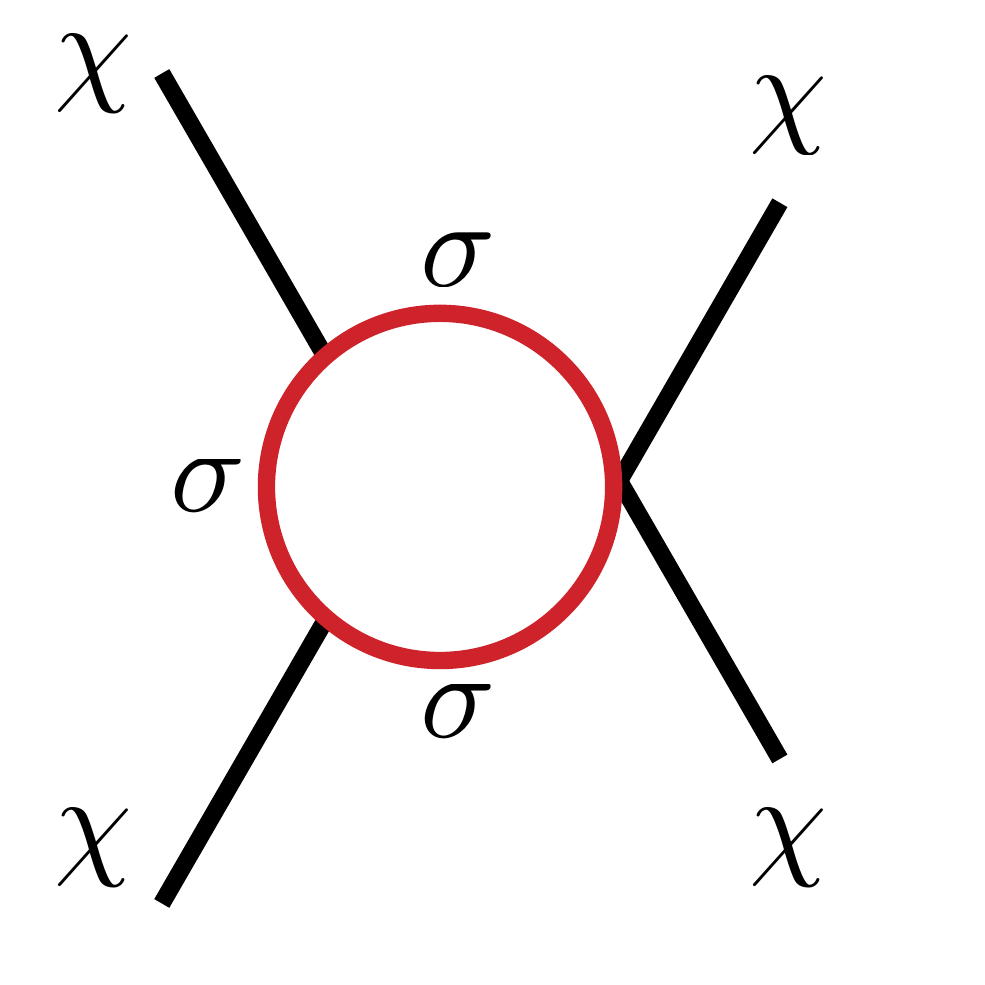} \quad 
	\includegraphics[width=4.5cm]{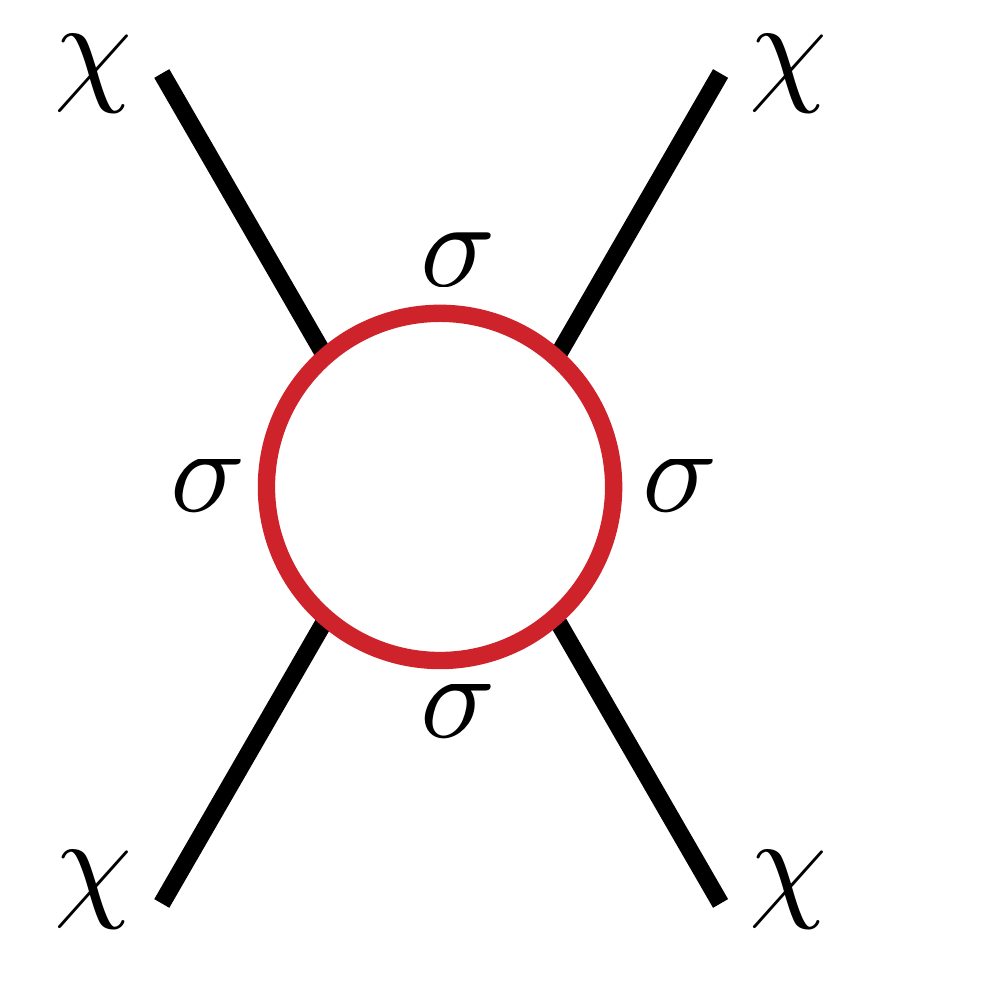} 
	\caption{The diagrams that contributes to $\langle \delta\chi \delta \chi  \delta \chi \delta \chi \rangle$.} \label{trispectrumchichichi}
\end{figure}

To calculate the trispectrum we consider the leading contributions from the 4-point diagrams plotted in Figure~\ref{trispectrumchichichi}. The contribution from the first diagram reads:
\begin{align} \nonumber
	& \langle \delta\chi_{\mathbf k_1} \delta\chi_{\mathbf k_2} \delta\chi_{\mathbf k_3} \delta\chi_{\mathbf k_4} \rangle'_{(2)} = \frac{1}{2} ( i \lambda)^2 \sum_{a,b=\pm} ab \int \frac{d \tau_1}{(-H\tau_1)^4}   \frac{d \tau_2}{(-H\tau_2)^4} G_a (k_1;\tau_2) G_a (k_2;\tau_2) G_b (k_3;\tau_1) \\ \label{fourpointcorrelationofchi}
	& \times G_b (k_4;\tau_1) \int \frac{d^3 \mathbf p}{(2\pi)^3}  \int \frac{d^3 \mathbf q}{(2\pi)^3} (2\pi)^3 \delta^{(3)} (\mathbf p+\mathbf q- \mathbf k_I) D_{ab} (p,\tau_1,\tau_2) D_{ab} (q,\tau_1,\tau_2) ~,
\end{align}
where we have defined $\mathbf k_I = \mathbf k_1+\mathbf k_2$. In the collapsed limit $k_1\simeq k_2\simeq k_3\simeq k_4\gg k_I$, it is 
\begin{align}\label{trispectruma}
	&  \langle \delta\chi_{\mathbf k_1} \delta\chi_{\mathbf k_2} \delta\chi_{\mathbf k_3} \delta\chi_{\mathbf k_4} \rangle'_{(2)} = \frac{-\lambda^2}{ 512 \pi^4 }  \frac{H^4 k_I^3}{ k_1^6 k_3^6} \bigg[ C_{\text{tri}}(\mu) \bigg(\frac{k_I^2 }{4 k_1 k_3}\bigg)^{-2i\mu}  +{\rm c.c.} \bigg] ~,
\end{align}
where
\begin{align}
	C_{\text{tri}}(\mu)=\mu^{-2} (2-i\mu)^2 \Gamma^2(i\mu) \Gamma^2(\frac{3}{2}-i\mu) \Gamma^2 (2-2 i \mu ) \Gamma(4i\mu-4)\sin(2i\mu\pi)\sin^2(i\mu\pi) ~.
\label{ctrispectrum}
\end{align}
\color{black}
Second and third Feynman diagram of Figure~\ref{trispectrumchichichi} is differed from the Feynman diagram on the leftmost by a factor of $2 \lambda \bar\chi^2/H^2$ and $2 \lambda^2 \bar\chi^4/H^4$, respectively. So the total contribution to the bispectrum is 
\begin{align}\label{trispectruma1}
	&  \langle \delta\chi_{\mathbf k_1} \delta\chi_{\mathbf k_2} \delta\chi_{\mathbf k_3} \delta\chi_{\mathbf k_4} \rangle' = \bigg(1+2\times \frac{ 2 \lambda \bar \chi^2 }{ H^2}+  \frac{ 2 \lambda^2 \bar \chi^4}{H^4}  \bigg)  \frac{-\lambda^2}{ 512 \pi^4 }  \frac{H^4 k_I^3}{ k_1^6 k_3^6} \bigg[ C_{\text{tri}}(\mu) \bigg(\frac{k_I^2 }{4 k_1 k_3}\bigg)^{-2i\mu}  +{\rm c.c.} \bigg] ~.
\end{align} 

Based on the same method, the first diagram in the derivative coupling case can be evaluated as
\begin{align} \nonumber
	& \langle \delta\chi_{\mathbf k_1} \delta\chi_{\mathbf k_2} \delta\chi_{\mathbf k_3} \delta\chi_{\mathbf k_4} \rangle'_{(2)} = \frac{1}{2} ( i \lambda_2)^2 \sum_{a,b=\pm} ab \int \frac{d \tau_1}{(-H\tau_1)^2}   \frac{d \tau_2}{(-H\tau_2)^2} G^{\prime}_a (k_1;\tau_2) G^{\prime}_a (k_2;\tau_2) G^{\prime}_b (k_3;\tau_1) \\ \label{fourpointcorrelationofchid}
	& \times G^{\prime}_b (k_4;\tau_1) \int \frac{d^3 \mathbf p}{(2\pi)^3}  \int \frac{d^3 \mathbf q}{(2\pi)^3} (2\pi)^3 \delta^{(3)} (\mathbf p+\mathbf q- \mathbf k_I) D_{ab} (p,\tau_1,\tau_2) D_{ab} (q,\tau_1,\tau_2) ~,
\end{align}
where we have defined $\mathbf k_I = \mathbf k_1+\mathbf k_2$. In the collapsed limit $k_1\simeq k_2\simeq k_3\simeq k_4\gg k_I$ and summing up all 3 leading diagrams, the result reads:
\begin{align}\label{trispectruma2}
	& \langle \delta\chi_{\mathbf k_1} \delta\chi_{\mathbf k_2} \delta\chi_{\mathbf k_3} \delta\chi_{\mathbf k_4} \rangle' = \bigg(1+2\times \frac{ 2 \lambda_2 \dot{\bar \chi}^2 }{ H^2}+  \frac{ 2 \lambda_2^2 \dot{\bar \chi}^4}{H^4}  \bigg)  \frac{\lambda_2^2}{ 2^{13} \pi^4 }  \frac{H^8 k_I^3}{ k_1^6 k_3^6} \bigg[ C_{\text{tri}}^d(\mu)  \bigg(\frac{k_I^2 }{4 k_1 k_3}\bigg)^{-2i\mu}  +{\rm c.c.} \bigg] ~.
\end{align} 
where the loop suppression factor for the derivative case
\begin{align}
	  C_{\text{tri}}^d(\mu)=\Gamma^2(i\mu) \Gamma^2(\frac{3}{2}-i\mu) \Gamma^2 (4-2 i \mu ) \Gamma(4i\mu-4) \sin(2i\mu\pi)\sin^2(i\mu\pi) ~.
\label{ctrispectrumd}
\end{align}

\section{Curvaton Scenario: NG Introduced by Light Field Decay}\label{curvaton}
In this section and the next section, we discuss the post-inflationary evolution of the isocurvature fluctuations. The discussion here is aligned with the so-called ``curvaton scenario" \cite{Enqvist:2001zp,Lyth:2001nq,Moroi:2001ct}
. The idea dates back to the earlier works \cite{Mollerach:1989hu,Linde:1996gt}. After inflation, the subsequent evolution consists of three stages: curvaton field starting oscillation, oscillation persisting for many Hubble times and curvaton decaying. The curvaton field can either dominate the energy density just before the decay or do not dominate during the whole process. Defining the ratio of the curvaton energy density $\rho_\chi$ and the radiation energy density $\rho_r$ to be $r_{\rm dec}$ at the time of curvaton decay, we have
\begin{align}
r_{\rm dec} \equiv \frac{3\rho_\chi}{4\rho_r+3\rho_{\chi}} \bigg|_{\rm decay} ~.
\end{align} 
Another parameter helpful in characterizing different scenarios is the ratio of the power spectrum of the primordial curvature perturbations generated by the $\chi$ field to that generated by the inflaton, which we denoted as
\begin{align}
	R \equiv \frac{P_\zeta^{(\chi)}}{P_\zeta^{(\phi)}}~. 
\end{align}

In this section we focus on the scenario where the power spectrum is mostly generated by $\chi$ ($R\gg 1$) and leave the scenario where the power spectrum is mostly generated by inflaton to the next section. The current measured value of primordial scalar power spectrum $P_\zeta^{(0)} = 2.1\times 10^{-9}$~\cite{Aghanim:2018eyx} demands that $\bar{\chi}$ has a fixed ratio with $H$:
\begin{equation}
\frac{\bar{\chi}}{H} \simeq  2.3\times 10^3~.
\end{equation}
We will simply fix this ratio in the rest of this section. Also, since by definition the curvaton would decay to radiation, it naturally satisfies the constraint that $\rho_\chi$ is negligible in the late time universe. We take the minimal form that $V(\chi)=\frac{m_\chi \chi^2}{2}$ here.

The post-inflationary evolution starts with a period when the curvaton field starts to oscillate around the minimum of the potential. During this time, the universe is dominated by radiation. The oscillation starts when the Hubble scale coincide with the mass of the light field $H \sim m$. 

The energy density of the $\chi$ field is
\begin{align}
	\rho_{\chi} (\mathbf x) = \rho_{\chi} + \delta \rho_{\chi} (\mathbf x) = \frac{1}{2} m_{\chi}^2 \chi^2 (\mathbf x) ~.
\end{align}
Based on the inhomogeneity we can define the density contrast and curvature perturbations as:
\begin{align}
	\delta &\equiv \frac{\delta \rho_{\chi}}{\rho_{\chi}}= 2 \frac{\delta \chi}{\chi}~. \\
	\zeta_{\chi} &=-H \frac{\delta \rho_{\chi}}{\dot{\rho}_{\chi}}=\frac{1}{3} \frac{\delta \rho_{\chi}}{\rho_{\chi}} =\frac{2}{3}\frac{\delta \chi}{\chi}=\frac{1}{3} \delta~.
\end{align}
Observation requires that \cite{Feldman:2000vk}$\bar \chi_* \gg H_*$ for curvature perturbations where $\chi_*$ and $H_*$ denotes the field value of $\chi$ and Hubble are evaluated at horizon crossing. 

The curvature perturbation in the radiation section is simply given by:
\begin{align} 
\zeta_{\mathrm{r}} &=-H \frac{\delta \rho_{\mathrm{r}}}{\dot{\rho}_{\mathrm{r}}}=\frac{1}{4} \frac{\delta \rho_{\mathrm{r}}}{\rho_{\mathrm{r}}} ~,
\end{align} 
The curvature perturbation is thus
\begin{align}\label{curvatureperturbation}
\zeta =-H \frac{\delta \rho}{\dot{\rho}}=-H \frac{\delta \rho_{\mathrm{r}}+\delta \rho_{\chi}}{\dot{\rho}_{\mathrm{r}}+\dot{\rho}_{\chi}} = \frac{4 \rho_r \zeta_r+ 3 \rho_\chi \zeta_\chi}{4 \rho_r+3 \rho_\chi}~.
\end{align}
The energy density of the $\chi$ field can either dominate or sub-dominate the energy density before the decay. In the following, we discuss these two cases separately. 
If $\chi$ field dominates the energy density before the decay ($r\sim 1$),  \eqref{curvatureperturbation} becomes
\begin{align}
	\zeta\simeq  \frac{r}{3 } \delta = \frac{2 r}{3} \frac{\delta \chi}{\chi } ~.
\end{align}

$\chi$ field sub-dominates the energy density before the decay. On the other hand, the case that the energy of the curvaton is subdominant compared with radiation $r\ll 1$, we will obtain
\begin{align}
\zeta \simeq \frac{r}{4}  \delta  ~.
\end{align} 
In this work we will focus on the first scenario where $r\sim 1$, as for the later case the calculations and results are qualitatively the same and the fact that the later case is strongly constrained by data.

\subsection{NG Signals}
The $\chi$ bispectrum and trispectrum are related to $\zeta$ trispectrum and bispectrum in the following way
\begin{align}
\langle \zeta_{\mathbf k_1} \zeta_{\mathbf k_2} \zeta_{\mathbf k_3} \rangle & = \bigg(\frac{2}{3} \frac{r}{\bar \chi} \bigg)^3 \langle \delta \chi_{\mathbf k_1} \delta  \chi_{\mathbf k_2} \delta  \chi_{\mathbf k_3} \rangle ~,\\
\langle \zeta_{\mathbf k_1} \zeta_{\mathbf k_2} \zeta_{\mathbf k_3} \zeta_{\mathbf k_4} \rangle & = \bigg( \frac{2}{3} \frac{r}{\bar \chi} \bigg)^4 \langle \delta \chi_{\mathbf k_1} \delta \chi_{\mathbf k_2} \delta \chi_{\mathbf k_3} \delta \chi_{\mathbf k_4} \rangle ~.
\end{align}
We define the shape function to be
\begin{align}
\left\langle\zeta_{\mathbf{k}_{1}} \zeta_{\mathbf{k}_{2}} \zeta_{\mathbf{k}_{3}}\right\rangle^{\prime} & \equiv (2 \pi)^{4} S\left(k_{1}, k_{2}, k_{3}\right) \frac{1}{\left(k_{1} k_{2} k_{3}\right)^{2}} P_{\zeta}^{(0) 2} ~,  \\
\left\langle \zeta_{\mathbf{k}_{1}} \zeta_{\mathbf{k}_{2}} \zeta_{\mathbf{k}_{3}} \zeta_{\mathbf{k}_{4}}\right\rangle^{\prime} & \equiv (2 \pi)^{6}  T\left(k_{1}, k_{2}, k_{3}, k_{4}\right) \frac{(k_1+k_2+k_3+k_4)^{3}}{\left(k_{1} k_{2} k_{3} k_{4}\right)^{3}} P_{\zeta}^{(0)3} ~.
\end{align}
So the shape functions are in the direct coupling case:
\begin{align}\nonumber
S(k_1,k_2,k_3) &= \frac{1}{(2\pi)^4 P^{(0)2}_\zeta}  \frac{8 r^3}{27\bar\chi^3}    \bigg(1+\frac{2\lambda \bar \chi^2}{H^2}\bigg)   \frac{-\lambda^2 \bar\chi H^2}{32 \pi^4}   \bigg[ C_{\text{bi}}(\mu) \bigg(\frac{k_3  }{2 k_1}\bigg)^{2-2i\mu}  +\text{c.c.} \bigg] ~, \\ \nonumber
T\left(k_{1}, k_{2}, k_{3}, k_{4}\right) &   = \frac{1}{(2\pi)^6 P^{(0)3}_{\zeta}}\frac{16 r^4}{81 \bar \chi^4} \bigg(1+ \frac{ 4 \lambda \bar \chi^2 }{ H^2}+  \frac{ 2 \lambda^2 \bar \chi^4}{H^4}  \bigg) \frac{-\lambda^2 H^4}{ 512 \pi^4 }  \frac{(k_1 k_3)^{3/2}}{(k_1 +k_3)^3}\times \\   
& \times
\bigg[C_{\text{tri}}(\mu) \bigg(\frac{k_I^2 }{4 k_1 k_3}\bigg)^{\frac{3}{2}-2i\mu}  +{\rm c.c.} \bigg] ~.
\label{eq:curvatonS}
\end{align}
Where the normalization factor is the primordial scalar power:
\begin{align}
\frac{R}{R+1}P^{(0)}_\zeta=P^{(\chi)}_\zeta=\frac{4r^2}{9\bar{\chi}^2}\bigg(\frac{H}{2\pi}\bigg)^2 \simeq 2.1\times 10^{-9} ~,
\end{align}
$C_{\text{bi}}(\mu)$ and $C_{\text{tri}}(\mu)$ are dimensionless functions of $\mu$, as defined in (\ref{cbispectrumd}) and (\ref{ctrispectrumd}). If we adopt the derivative coupling, following a similar approach the shape functions becomes:
\begin{align}
S(k_1,k_2,k_3) &= \frac{1}{(2\pi)^4 P^{(0)2}_\zeta}  \frac{8 r^3}{27\bar\chi^3}\bigg(1+\frac{2\lambda_2 \dot{\bar \chi}^2}{H^2}\bigg)    \frac{-\lambda_2^2 \dot{\bar\chi} H^5}{64 \pi^4}   \bigg[ C_{\text{bi}}^d (\mu)  \bigg(\frac{k_3  }{2 k_1}\bigg)^{2-2i\mu}  +\text{c.c.} \bigg] ~,
\end{align}
\begin{align}\nonumber
T\left(k_{1}, k_{2}, k_{3}, k_{4}\right) &= \frac{1}{(2\pi)^6 P^{(0)3}_{\zeta}}\frac{16 r^4}{81 \bar \chi^4}\bigg(1+\frac{ 4 \lambda_2 \dot{\bar \chi}^2 }{ H^2}+  \frac{ 2 \lambda_2^2 \dot{\bar \chi}^4}{H^4}  \bigg)  \frac{\lambda_2^2 H^8}{ 2^{13} \pi^4 } \frac{(k_1 k_3)^{3/2}}{(k_1 +k_3)^3}\times\\ & \times \bigg[ C_{\text{tri}}^d(\mu)  \bigg(\frac{k_I^2 }{4 k_1 k_3}\bigg)^{\frac{3}{2}-2i\mu} +{\rm c.c.} \bigg] ~,
\label{eq:curvatonDS}
\end{align}

\begin{figure}
\centering
\includegraphics[height=6cm]{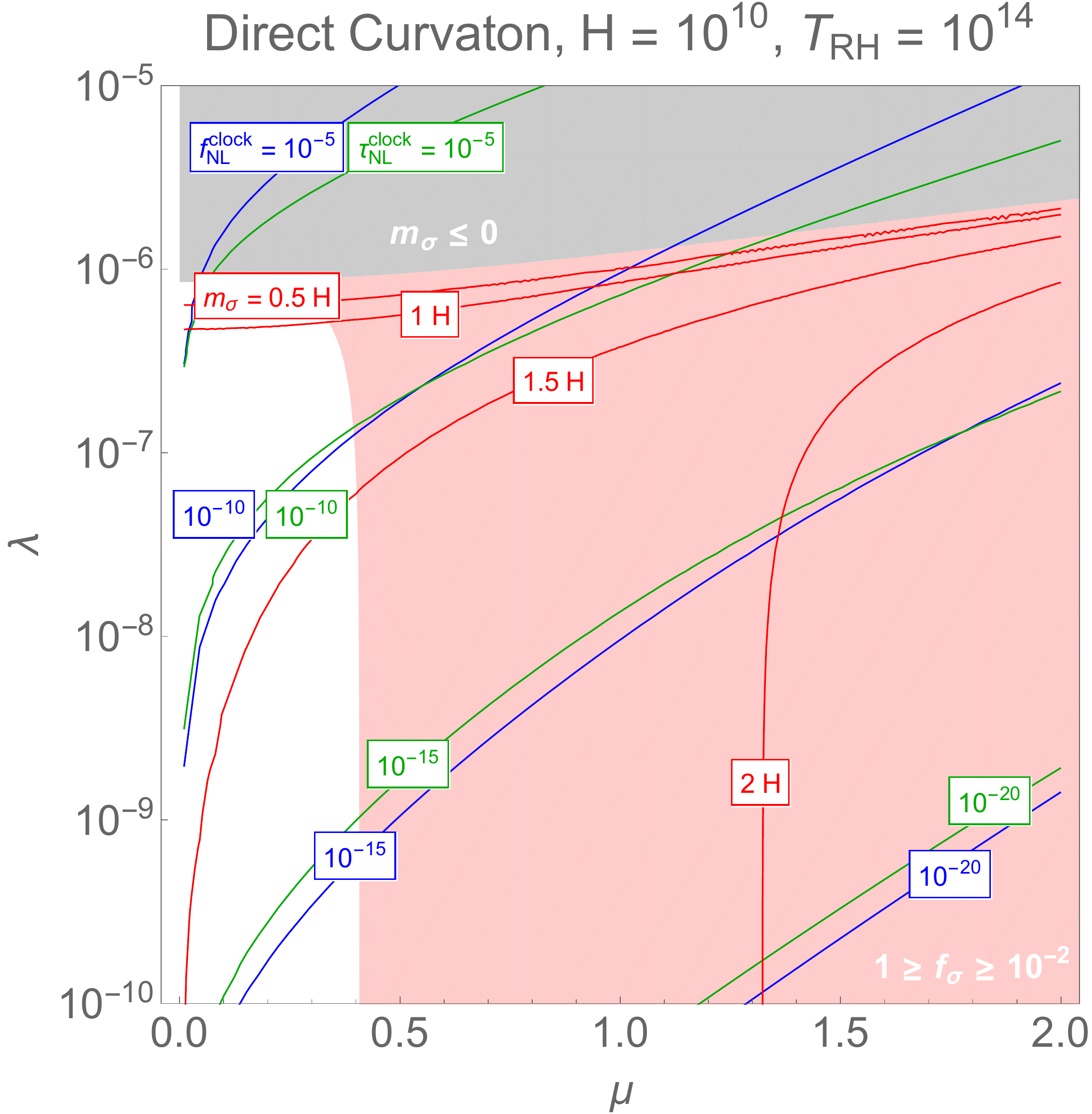}
\includegraphics[height=6cm]{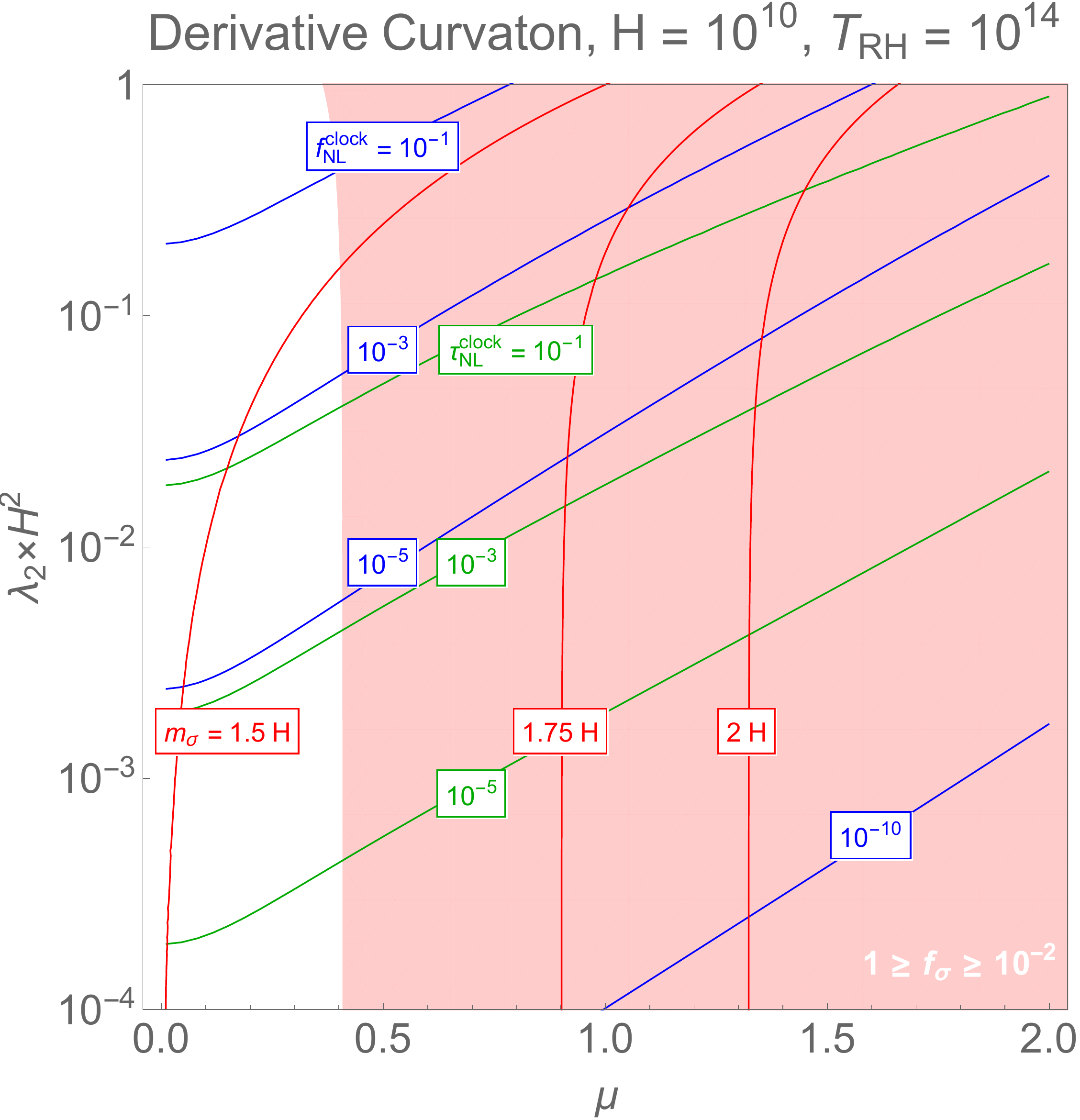}
\caption{Numerical benchmarks of the curvaton scenario, $H$ and $T_{\rm RH}$ in the unit of GeV. The gray region corresponds to when the bare $m_\sigma$ becomes negative, and the red contours stand for $m_\sigma$ in the unit of $H$. The blue(green) contour stands for bispectrum(trispectrum). The red region provides enough DM this way( $1>f_\sigma>1\%$ in the range of $10^{9}>A>10^{4}$). {\textbf Left:} Direct coupling. {\textbf Right:} Derivative coupling.}
\label{fig:curvatonDM}
\end{figure}

In Fig.~\ref{fig:curvatonDM} we show two typical numerical parameter space for this scenario. The parameter choice are labeled on each plot. We simply take $r=1$ to fulfill our expectation that $r\sim \mathcal{O}(1)$. In fact, the size of $r$ doesn't affect the curvature NG clock signal significantly as it only depends on $r^3$ rather than exponentially. 
In both cases $R$ is sufficiently larger than 1, consistent with our assumption for curvaton case. In most cases, both the size of bispectrum $S$ and scaled DM relic density $f_\sigma$ are sensitive to $\mu$. The model thus prefers smaller $\mu$ that produces enough $\sigma$ DM and large $S$.

\subsection{Observational Constraints}

For the curvaton-like case, the power spectrum is dominated by curvaton decays. For formally one can define the ratio
\begin{equation}
\label{eq:Rdefinition}
R\equiv \frac{\Pcal^\chi_\zeta}{\Pcal^\phi_\zeta} \simeq \frac{8\epsilon}{9} \bigg(\frac{\Mpl}{\bar{\chi}} \bigg)^2 r_{\rm dec}^2~.
\end{equation} 
According to previous studies \cite{Enqvist:2013paa}, the curvaton itself can introduce a non-trivial $f_{\rm NL}$: 
\begin{equation}
\label{eq:curvatonfnl}
f_{\rm NL} = \frac{5}{6(1+R)^2} \bigg[\frac{1}{2\epsilon} \bigg( 1-\frac{\eta_\phi}{2\epsilon}\bigg) + R^2 \bigg( \frac{3- 4r_{\rm dec} -2r_{\rm dec}^2}{2 r_{\rm dec}}  \bigg) \bigg]~.
\end{equation}
As long as $R\gg 1$ as preferred by data, the observational constraint that $f_{\rm NL}< \mathcal{O}(10)$ then requires $r$ to be of $\mathcal{O}(1)$. 

Another important constraint comes from matter-photon isocurvature, which depends on the perturbation differences between the rest of CDM ($c$, other than $\sigma$), radiation ($\gamma$) and baryon (b). We define the gauge invariant entropy fluctuations $S_{x\gamma} \equiv 3 (\zeta_x - \zeta_\gamma )$, where $x$ denotes $b$ or $c$. If $\sigma$ is the dominant component of CDM, since it is created before curvaton decays, its perturbation imprints the information $\Pcal_\phi$ instead of $\Pcal_\chi$. In this model, the production of DM may even correlate to $\delta\chi$ negatively as $\bar{\chi}$ raises DM effective mass, which may make things even worse. However, as $\frac{\partial \ln\rho_{\sigma}}{\partial{\bar{\chi}}}\ll 1$ we can take only the leading contribution to isocurvature, which is not related to $\lambda$.

The current measurement strongly disfavor cases that all CDM are created before curvaton $\chi$ decay~\cite{Smith:2015bln,Aghanim:2018eyx} as this creates a large $S_{m\gamma}/\zeta\gtrsim \mathcal{O}(1)$. Therefore in the curvaton scenario we only consider if $\sigma$ is a subdominant part of CDM, while the rest of CDM is created after or by $\chi$ decay:
\begin{equation}
f_{\sigma}\equiv \frac{\Omega_{\sigma}}{\Omega_{\rm CDM}}< 1~.
\end{equation}
Aside from when CDM generation, baryon number $B$ creation can also important, assuming they are created in the radiation, if it is created before/by/after $\chi$ decay makes difference. On the other hand, the lepton number $L$, as long as it is not created by curvaton $\chi$ decay, makes little difference in our calculation ($R_\nu=0$). For more details, see~\cite{Smith:2015bln}. 

\begin{figure}[h!]
\centering
\includegraphics[scale=0.3]{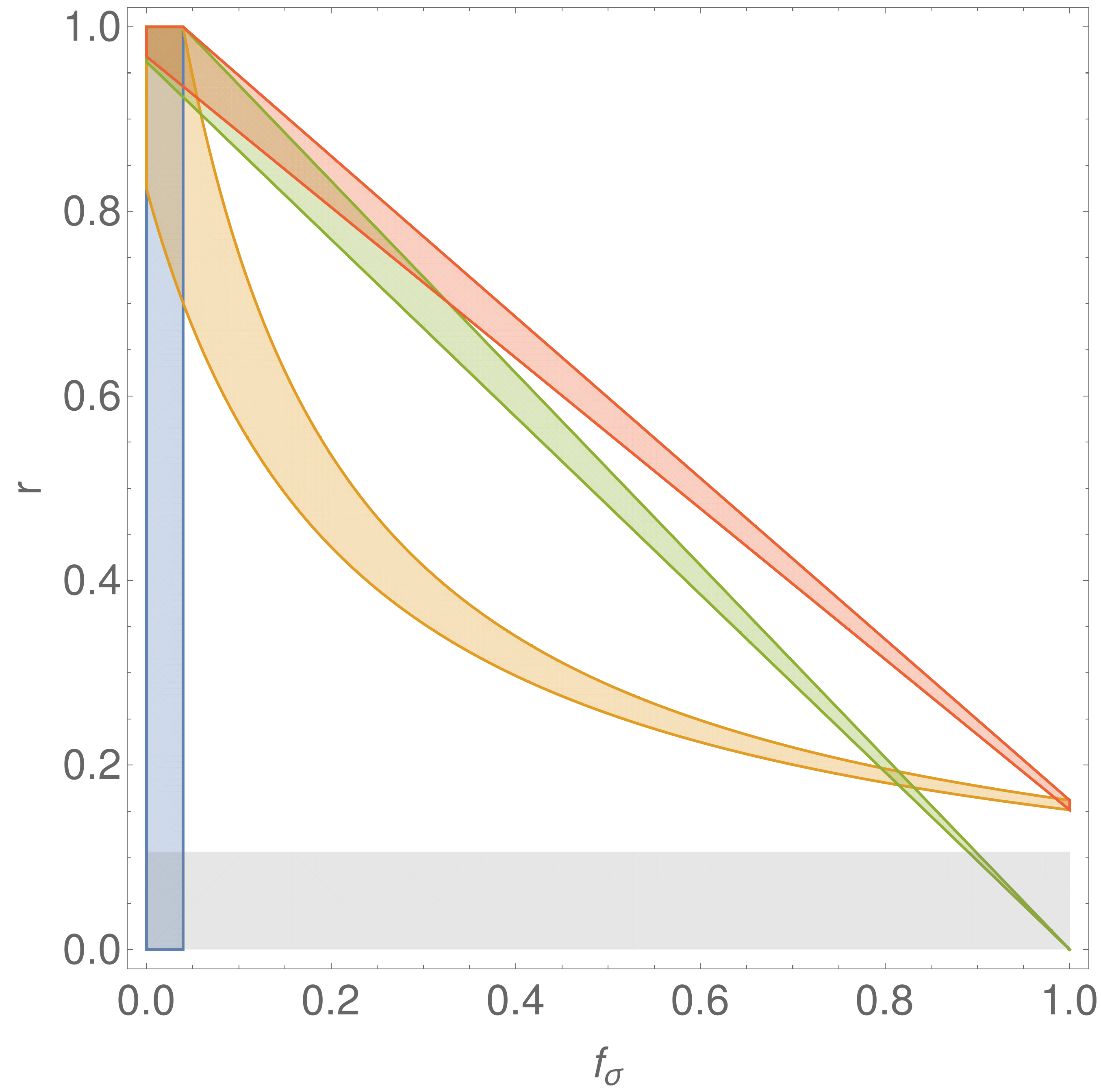}
\caption{Possible range of $r$ and $f_\sigma$ for 4 different scenarios, with the isocurvature constraint. Blue, green, orange and red regions stand for $b_{after},c_{after}$ /$b_{by},c_{after}$/  $b_{after},c_{by}$ and $b_{by},c_{by}$ respectively. The gray shade is vetoed as $r$ becomes too low and introduces too large a $f_{\rm NL}^{\rm local}>10$.}
\label{fig:rfscan}
\end{figure}

Following previous discussion, definition of the CDM fraction $f_\sigma$, and calculation in~\cite{Smith:2015bln}, one can calculate $S_{m\gamma}$ different cases. For example, if both the rest of CDM and $B$ are created after $\chi$ decay ($b_{after},c_{after}$):
\begin{equation}
\frac{S_{c\gamma}}{\zeta} = -3 f_\sigma,~\frac{S_{b\gamma}}{\zeta}=0,~\frac{S_{m\gamma}}{\zeta}=-3 f_\sigma \frac{\Omega_{\rm CDM}}{\Omega_m}~.
\end{equation}
Similarly, we have the case $b_{by},c_{after}$:
\begin{equation}-
\frac{S_{c\gamma}}{\zeta} = 3 f_\sigma,~\frac{S_{b\gamma}}{\zeta}=3(r^{-1}-1),~\frac{S_{m\gamma}}{\zeta}=\frac{3 \Omega_b[1+r(f_\sigma-1)]}{r \Omega_m} -3 f_\sigma ~,
\end{equation}
$b_{after},c_{by}$
\begin{equation}
\frac{S_{c\gamma}}{\zeta} = 3(r^{-1}-1)(1- f_\sigma) - 3 f_\sigma ,~\frac{S_{b\gamma}}{\zeta}=0,~\frac{S_{m\gamma}}{\zeta}=-\frac{3 \Omega_{\rm CDM}(r-1+f_\sigma)}{r \Omega_m} ~,
\end{equation}
and $b_{by},c_{by}$
\begin{equation}
\frac{S_{c\gamma}}{\zeta} = 3(r^{-1}-1)(1- f_\sigma) - 3 f_\sigma,~\frac{S_{b\gamma}}{\zeta}=3(r^{-1}-1),~\frac{S_{m\gamma}}{\zeta}=3\frac{\Omega_m(1-r) -\Omega_{\rm CDM} f_\sigma}{r \Omega_m} ~.
\end{equation}
 Notice that this immediately rules out the case that $f_\sigma$ is close to 1 for all different scenarios. Together with current data constraints $|S_{m\gamma}/\zeta|\lesssim \mathcal{O}(0.1)$, we can work out the possible range of $r$ and $f_\sigma$ based on the arguments is shown in Fig.~\ref{fig:rfscan}.

We also need to take the tri-spectrum $\tau_{NL}$ constraint into consideration. According to~\cite{Enqvist:2013paa}:
\begin{align}
\tau_{NL} = \bigg( \frac{1}{1+R} \bigg)^3 \bigg[ \frac{1}{4\epsilon^2}\bigg(1-\frac{\eta_\phi}{2\epsilon}\bigg)^2 + R^3 \bigg( \frac{3- 4r -2r^2}{2 r}  \bigg) \bigg]
\simeq \frac{R}{1+R} \bigg(\frac{6}{5} f_{NL} \bigg)^2~,
\end{align}
\begin{align}
g_{NL} = & \bigg( \frac{1}{1+R} \bigg)^3 \bigg[ 2\epsilon \bigg( - \frac{\xi-\phi^2}{\epsilon} + 2\frac{\eta_\phi^2}{\epsilon}\bigg) + R^3 \bigg( \frac{6r^3+20r+r-18}{2 r}  \bigg) \bigg] \\ \nonumber
\simeq &  \bigg(\frac{R}{1+R}\bigg)^3 \frac{6r^3+20r+r-18}{2 r}  .
\end{align}
The current constraint on $g_{\rm NL}$ is of $\mathcal{O}(10^4)$~\cite{Akrami:2019izv} and $\tau_{NL}<2800$~\cite{Ade:2013ydc}. With the assumption that $R\gg 1$, the constraints from the trispectrum measurements cannot provide meaningful information about this scenario as long as $r$ is of $\mathcal{O}(1)$.

\section{DM modulation Scenario: NG Introduced by DM Production}\label{spectator}
In this section, we will turn to another limit that $\chi$ barely leaves any curvature perturbation observable, in contrast to the curvaton case. Or equivalently we are discussing the $R\to 0$, $r\to 0$ limit using the notation we use in the curvaton case. As inflaton dominates both the curvature perturbation power spectrum and the energy density when $\chi$ decays, the NG imprints left by $\chi$ will be suppressed by $r^n$ and would not introduce visible $f_{\rm NL}$ constraints according to~\eqref{eq:curvatonfnl}. Nevertheless, in our model where $\chi$ directly couples to the dark matter $\sigma$, the information of $\chi$ primordial perturbation will also introduce isocurvature fluctuations in DM, which may become significant. The fluctuation in $\chi$ during inflation creates NG in terms matter-radiation isocurvature. Certainly, such a model must also be constrained by current observation of isocurvature power spectrum. In the follow section we will show that in order to get a large NG effect in terms of isocurvature, $f_\sigma$ need to be larger and eventually taken to be 1.

\subsection{Review of this Scenario}
 According to~\cite{Li:2019ves} and discussions in Sec.~\ref{DMproduction}, the heavy DM density produced by gravity is proportional to $\mu^\alpha e^{-2 \pi \mu}$. For superhorizon modes, the DM relic density before $\chi$ decays will pick up an non-zero fluctuation term due to the contribution from $\chi$ fluctuation:
\begin{equation}
\rho_\sigma = \bar{\rho}_\sigma e^{3(\zeta_\sigma-\zeta)}~,
\end{equation}
since during this era we have $\rho_\sigma\ll\rho_\chi\ll\rho_r$ and $\zeta_r \simeq \zeta_\phi$. Then, up to the leading order, we can expand the expression and get the following expression.
\begin{align}
	\zeta_\sigma-\zeta_\phi = \frac{1}{3} \frac{\delta \rho_\sigma}{\bar{\rho}_\sigma}  = \frac{1}{3} \frac{\frac{\partial \rho_\sigma}{\partial \dot{\chi}} \delta \dot{\chi}}{\bar{\rho}_\sigma } = \frac{1}{3} \delta \dot{\chi} \frac{\partial \ln \rho_\sigma}{\partial \dot{\chi}} = \frac{1}{3} \delta \dot{\chi} \frac{\partial \ln \rho_\sigma}{\partial \dot{\bar \chi}}~.
\end{align}
Note that inflaton also contributes to the curvature perturbation, however, it gives subdominant contributions. For more details, see \cite{Fonseca:2012cj}.

Since $\zeta_\phi$ and $\zeta_{\chi/\sigma}$ is uncorrelated, the main contributor to the NG of $\sigma$ density mostly stems from the isocurvature mode. This means the three-point function
\begin{align}
\label{eq:modbispectrum}
\lran{\frac{\delta\rho_{\sigma\mathbf k_1}}{\rho_\sigma }\frac{\delta\rho_{\sigma\mathbf k_2}}{\rho_\sigma }\frac{\delta\rho_{\sigma\mathbf k_3}}{\rho_\sigma }} &= \lran{\delta\chi_{\mathbf k_1} \delta\chi_{\mathbf k_2} \delta\chi_{\mathbf k_3}} \bigg(
\frac{\partial \ln \rho_\sigma}{\partial \bar{\chi}}
\bigg)^3\\\nonumber
= & \lran{\zeta_{\chi\mathbf{k}_{1}} \zeta_{\chi\mathbf{k}_{2}} \zeta_{\chi\mathbf{k}_{3}}} \bigg(\frac{3\bar{\chi} }{2r}
\frac{\partial \ln \rho_\sigma}{\partial \bar{\chi}}
\bigg)^3 \simeq  \lran{\zeta_{\chi\mathbf{k}_{1}} \zeta_{\chi\mathbf{k}_{2}} \zeta_{\chi\mathbf{k}_{3}}} \bigg(\frac{-3 \pi \lambda \bar{\chi}^2}{2r H^2 \mu} \bigg)^3~,
\end{align}
and for derivative coupling the above relation changes to:
\begin{align}
& \lran{\frac{\delta\rho_{\sigma\mathbf k_1}}{\rho_\sigma }\frac{\delta\rho_{\sigma\mathbf k_2}}{\rho_\sigma }\frac{\delta\rho_{\sigma\mathbf k_3}}{\rho_\sigma }} = \lran{\delta\dot{\chi}_{\mathbf k_1} \delta\dot{\chi}_{\mathbf k_2} \delta\dot{\chi}_{\mathbf k_3}} \bigg( \frac{\partial \ln \rho_\sigma}{\partial \dot{\bar{\chi}}} \bigg)^3
\simeq \lran{\delta{\chi}_{\mathbf k_1} \delta{\chi}_{\mathbf k_2} \delta{\chi}_{\mathbf k_3}} \bigg(\frac{-m^2_\chi}{3H}\bigg)^3 \bigg( \frac{\partial \ln \rho_\sigma}{\partial \dot{\bar{\chi}}} \bigg)^3
\\\nonumber
& =  \lran{\zeta_{\chi\mathbf{k}_{1}} \zeta_{\chi\mathbf{k}_{2}} \zeta_{\chi\mathbf{k}_{3}}} \bigg(\frac{3\bar{\chi} }{2 r} \bigg)^3 \bigg(\frac{-m^2_\chi}{3H}\bigg)^3 \bigg( \frac{\partial \ln \rho_\sigma}{\partial \dot{\bar{\chi}}} \bigg)^3
\simeq  \lran{\zeta_{\chi\mathbf{k}_{1}} \zeta_{\chi\mathbf{k}_{2}} \zeta_{\chi\mathbf{k}_{3}}}\bigg(\frac{-m^2_\chi\bar{\chi}}{2Hr}\frac{-\pi \lambda_2 \dot{\bar{\chi}}}{ H^2 \mu}\bigg)^3~,
\label{eq:modbispectrum}
\end{align}
where in the last relation we assume $\alpha\simeq 0$ in \eqref{eq:DMdensity3} for simplicity. The term grabs a negative sign compared to the $\chi$ induced clock signal. The calculation of the clock signal is then similar to the case in Sec.~\ref{curvaton}. However, to correctly compare the NG in isocurvature with the Gaussian part, the normalization factor changed to primordial isocurvature perturbation at large scale $P_I^{(0)}$ deduced from the same mechanism instead. The current experiment only gives an upper limit of the size. In this mechanism, the primordial isocurvature perturbation can be introduced by modulated $\sigma$ dark matter production similar to in, which will be discussed below.

\begin{figure}[h!]
\centering
\includegraphics[height=6cm]{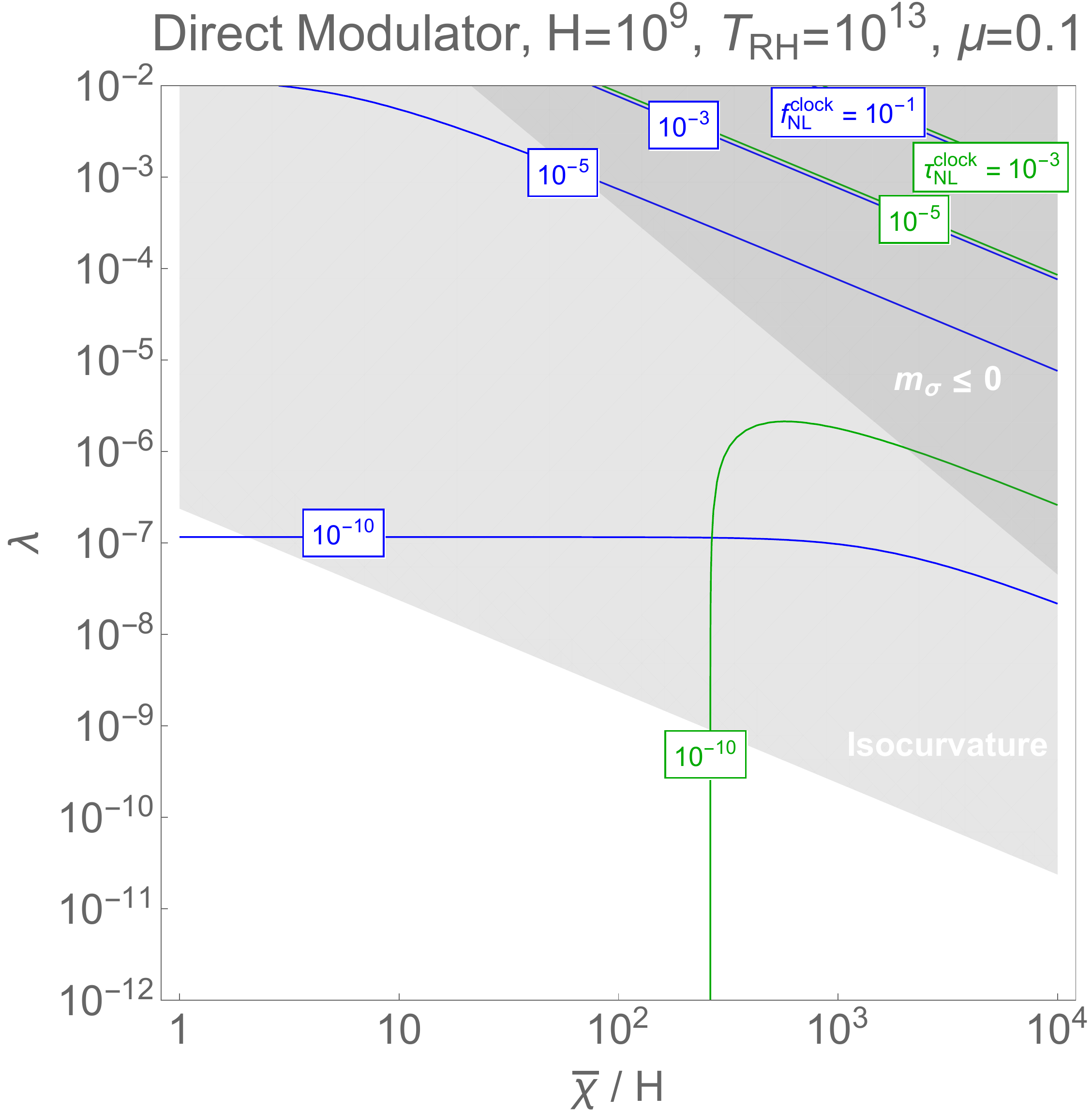}
\includegraphics[height=6cm]{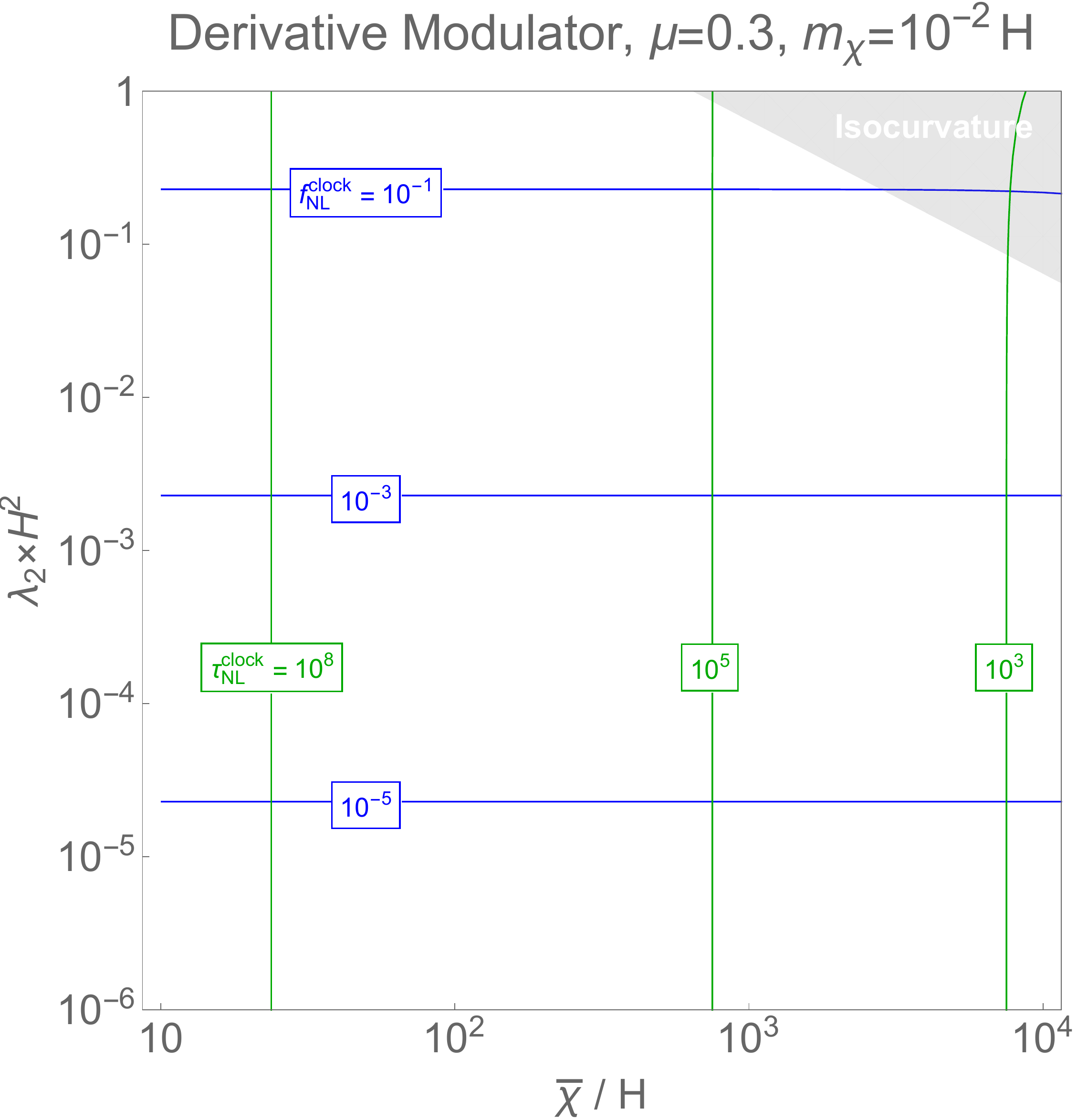}
\caption{The parameter scan for the modulator case, $H$ and $T_{\rm RH}$ in the unit of GeV. Blue and green contours still stand for the log of bi/trispectrum. The light gray shaded region is prohibited by the isocurvature perturbation observation bound. The dark gray shaded region is where $m_\sigma$ turns negative with given parameter. In both cases all allowed (blank) region provides enough $\sigma$ DM ($f_\sigma$=1) with a proper choice of parameter $A$ between $10^4$ and $10^9$.}
\label{fig:mod_1}
\end{figure}

Now we turn to estimate the primordial isocurvature perturbation power spectrum introduced by $\chi$-$\sigma$ interaction at large scales, assuming no other source of isocurvature. Similar to the curvaton case but with much smaller $R$ and $r$, the $\zeta_\gamma$ mainly follows inflaton fluctuation:
\begin{equation}
\zeta_\gamma \simeq  \zeta_\phi~,
\end{equation}
while $\zeta_c$ and $\zeta_b$ is assumed to be created by radiation since $\chi$ vanishes. This is corresponding to the $c_{\rm after}, b_{\rm after}$ scenario, leaving $\zeta_b \simeq \zeta_\gamma$ and$
\zeta_{\rm DM}=f_\sigma \zeta_\sigma + (1-f_\sigma)\zeta_\gamma~.$ Resulting in the total matter non-adiabatic perturbation
\begin{equation}
S_{m\gamma} = 3\frac{\Omega_{\rm CDM} f_\sigma}{\Omega_m}(\zeta_\sigma-\zeta_\gamma) \simeq \frac{\Omega_{\rm CDM} f_\sigma}{\Omega_m} \frac{\partial \ln \rho_\sigma}{\partial \bar \chi}\delta\chi \approx \frac{f_\sigma \lambda \bar{\chi}}{2 H \mu}.
\label{eq:direct}
\end{equation}
In the last relation we use the approximation that $\delta \chi \approx H/2\pi$ and the fact that $\Omega_b$ is only a small fraction of $\Omega_m$. According to the current isocurvature constraints~\cite{Akrami:2018odb}, the value of $\lambda \chi/H\mu $ must be small enough ($\lesssim \mathcal{O}(10^{-5})$). Similarly, we can write down the $S_{m\gamma}$ for the derivative coupling case:
\begin{equation}
S_{m\gamma} = 3\frac{\Omega_{\rm CDM} f_\sigma}{\Omega_m}(\zeta_\sigma-\zeta_\gamma) \simeq \frac{\Omega_{\rm CDM} f_\sigma}{\Omega_m} \frac{\partial \ln \rho_\sigma}{\partial \dot{\bar \chi}}\delta\dot{\chi} \approx \frac{f_\sigma \lambda_2 \dot{\bar{\chi}}^2}{2H \bar{\chi} \mu}~,
\label{eq:deriso}
\end{equation}
which is solvable once we impose the relation between $\dot{\bar{\chi}}$ and $\bar{\chi}$ such as in Eq.~\ref{eq:chislowroll}. The constraint from primordial isocurvature is shown in Figure~\ref{fig:mod_1} as the gray shades. It is obvious from the form of Eq.~\ref{eq:deriso} that the dependence on $\dot{\bar{\chi}}^2/\bar{\chi}H \propto m_\chi^4$ is small since $\chi$ is a light field. With a reasonable choice of $m_\chi$, such as $0.01H$ we chose here, the constraint from primordial isocurvature are much weaker, allowing larger NG signals. 

Since in the modulator scenario the light field $\chi$ no longer generates the primordial curvature spectrum, the ratio $\bar{\chi}/H$ is no longer fixed and hence becomes a free parameter. In this work we are particularly interested in the region where $\bar{\chi}/H$ is larger than 1 to use Eq.~\ref{eq:chislowroll} directly. Therefore the parameter space we shall have an additional dimension compared to the curvaton case. If we fix $\mu$ to be $\lesssim 1$ which is required to have enough NG, we can scan through the parameter space. We show the result of scans in both direct coupling and derivative coupling in Fig.~\ref{fig:mod_1}. In both panels, the whole allowed regions can provide enough $\sigma$ DM ($f_\sigma=1$) with a proper choice of parameter $A$ between $10^4$ and $10^9$. For direct coupling case it would be very hard to give a large NG signal without extreme fine tunings. The reason is due to the fact that only 2 parameters: the coupling $\lambda$ and the ratio $\bar{\chi}/H$ controls the strength of NG signals. Instead, for the derivative coupling case, a much larger NG signal can be achieved without much fine tuning ($\mu\gtrsim 0.1$). 

\section{Conclusion and Outlook}\label{summary} 

We propose a mechanism that a light spectator field can leave nontrivial observation signals via its interaction between gravitationally produced heavy DM. The minimal model consists of an inflaton $\phi$, a massive field $\sigma$ and a light field $\chi$. The conclusion is largely inflation model independent as all non-trivial interactions happens inside the $\sigma$-$\chi$ sector and the inflaton $\phi$ only acts as a background field. Specifically we consider two types of well-motivated couplings, direct coupling and derivative one. For both type of couplings the primordial three-point and four-point correlation functions of the light field $\chi$ mediated by the heavy DM field $\sigma$ are calculated. The $Z_2$ symmetry that protects $\sigma$ demands that such interactions can only present at loop level, indicating suppressed NG signals. However, large enough signal can still appear with proper $m_\sigma$ and derivative couplings.

The first case we propose is the well-known curvaton scenario, where the curvature perturbation we observe today are dominated by $\chi$. The current matter isocurvature observations strongly constraints the $\sigma$ heavy DM relic density. Therefore in this scenario we consider the curvature NG introduced by $\chi$ after it decays to radiation. The curvaton-heavy DM derivative interaction still leads to measurable clock signals while the direct coupling cannot provide enough NG.

In another scenario in which $\chi$ energy density and power spectrum is always sub-dominant, it is difficult to measure $\chi$ properties via curvature perturbations. However, the $\sigma$ heavy DM production is sensitive to $\chi$ field values during the inflationary era. Consequently, the matter isocurvature mode keeps the imprints of the primordial $\chi$ field configuration. The result of both types of couplings are shown. 
We will leave other intermediate cases to future studies.

In this work, we have focused on relatively simple cases. There can be many possibilities to further enhance the signal, including symmetry breaking, chemical potential, multiple dark matter particles with different masses, and so on. It would be interesting to explore these possibilities.

\section*{Acknowledgements}
We would like to thank Tomohiro Nakama for useful discussions. SL and YW were supported in part by GRF Grants 16301917, 16304418 and 16303819 from the Research Grants Council of Hong Kong. The work of SZ was supported in part by the Swedish Research Council under grants number 2015-05333 and 2018-03803.

\appendix
\section{Details of the Computation of the Loop Diagram}
\label{loopdetails}
In this section, we present the details of the derivation of the cosmological collider signal of \eqref{trispectruma1}, \eqref{trispectruma2}, \eqref{eq:curvatonS} and \eqref{eq:curvatonDS}. We used the Schwinger-Keldysh formalism \cite{Chen:2017ryl}. Alternatively, one can use the in-in formalism \cite{Weinberg:2005vy,Chen:2010xka,Wang:2013eqj}. 
   
Every diagram we needed to calculate has the same structure for loop integral
\begin{align}
	\mathcal{I}(\mathbf k_I,\tau_1,\tau_2)=\int\frac{d^3\mathbf p}{(2\pi)^3}D_{ba}(p,\tau_1,\tau_2)D_{ba}(|\mathbf k_I-p|,\tau_1,\tau_2) ~,
\end{align}
this can also be written as
\begin{align}
	\int d^3 \mathbf X e^{-i \mathbf k_I\cdot X }\frac{1}{2}\langle \sigma^2(x_1) \sigma^2(x_2) \rangle_{ba} ~,
\end{align}
in position space, where $x_1=(\tau_1,\mathbf x_1)$, $x_2=(\tau_2,\mathbf x_2)$, and $\mathbf X=\mathbf x_2-\mathbf x_1$. As we are only interested in non-local part, we can drop $ab$ indices here\cite{Chen:2016hrz}. 
\begin{align}
	\mathcal{I}( k_I;\tau_1,\tau_2)=	\int d^3 \mathbf X e^{-i \mathbf k_I\cdot X }D^2(x_1,x_2) ~.
\end{align}
One method to get the propagators in position space is to transform from de Sitter space $dS_D$ to $S^D$ sphere, with the help of spherical harmonics. The propagator for real scalar field is
\begin{align}
	D(x_1,x_2)=\frac{H^{D-2}}{(4\pi)^{D/2}}\frac{\Gamma(d/2-i\mu)\Gamma(d/2+i\mu)}{\Gamma(D/2)}\ _2F_1\left(\frac{d}{2}-i\mu,\frac{d}{2}+i\mu ;\frac{D}{2},\frac{1+Z_{x_1,x_2}}{2}\right) ~,
\end{align}
where $Z_{x_1,x_2}$ is the sphere imbedding distance $
	Z(x_1,x_2)=1-\frac{|\mathbf x_1-\mathbf x_2|^2-(\tau_1-\tau_2)^2}{2\tau_1\tau_2}
$, $\mu=\sqrt{(m_\sigma/H)^2-(d/2)^2}$, and $d\equiv D-1=3$ in our case. The propagator can be expanded at late time limit
\begin{align}
	D(x_1,x_2)\xrightarrow{\tau_1,\tau_2\rightarrow 0}\frac{H^2}{4\pi^{5/2}}\left[\Gamma(i\mu)\Gamma(\frac{3}{2}-i\mu)\left(\frac{\tau_1\tau_2}{X^2}\right)^{3/2-i\mu}+(\mu\rightarrow -\mu)\right] ~,
\end{align}
we can get
\begin{align}\nonumber
	\mathcal{I}(k_I;\tau_1,\tau_2)\xrightarrow{\tau_1,\tau_2\rightarrow 0}&\frac{H^4}{4\pi^4}\left[\Gamma(i\mu)^2\Gamma(\frac{3}{2}-i\mu)^2\Gamma(4i\mu-4)\sin(2\pi i\mu)(\tau_1\tau_2)^{3-2i\mu}k_I^{3-4i\mu}+(\mu\rightarrow -\mu)\right]\\
	=&\frac{H^4}{4\pi^4}\left[I(\mu)(\tau_1\tau_2)^{3-2i\mu}k_I^{3-4i\mu}+(\mu\rightarrow -\mu)\right] ~,
\end{align}
where $I(\mu)$ is defined as
\begin{align}
	I(\mu)=\Gamma(i\mu)^2\Gamma(\frac{3}{2}-i\mu)^2\Gamma(4i\mu-4)\sin(2\pi i\mu) ~.
\end{align}

\paragraph{Three-point function with $\mathbf \lambda$:}
The relevant part is the time integral. The three-point function \eqref{threepointcorrelationofchi} for $\lambda$ coupling case
\begin{align}\nonumber
	\langle \delta\chi_{\mathbf k_1} \delta\chi_{\mathbf k_2} \delta\chi_{\mathbf k_3} \rangle'_{(2)} =& \frac{1}{2} (i \lambda)^2 \bar\chi \sum_{a,b=\pm} ab \int \frac{d \tau_1}{(H\tau_1)^4}   \frac{d \tau_2}{(H\tau_2)^4} G_a (k_1;\tau_2) G_a (k_2;\tau_2) G_b (k_3;\tau_1)
\mathcal{I}(k_3;\tau_1,\tau_2)\\ \nonumber
	\xrightarrow{\tau_1,\tau_2\rightarrow 0}\simeq & \frac{- \lambda^2 \bar\chi}{2}H^{-8}  \frac{H^6}{8 k^3_1 k^3_2 k^3_3}\frac{H^4}{4\pi^4} \int^0_{-\infty}  d \tau_1 \int^0_{-\infty} d \tau_2\ \tau_1^{-4} \tau_2^{-4}  \sum_{a,b=\pm} ab(1-i b k_3\tau_1)\times\\ 
\times & (1-i a k_1\tau_2)(1-i a k_2\tau_2)e^{i a k_{12}\tau_2}e^{i b k_3\tau_1}I(\mu)(\tau_1\tau_2)^{3-2i\mu}k_3^{3-4i\mu}+(\mu\rightarrow -\mu) ~,
\end{align}
where $k_{12}=k_1+k_2$, 
\begin{align}\nonumber
	\langle \delta\chi_{\mathbf k_1} \delta\chi_{\mathbf k_2} \delta\chi_{\mathbf k_3} \rangle'_{(2)}\simeq & \frac{- \lambda^2 \bar\chi}{2 \pi^4} \frac{ H^{2}}{32 k^3_1 k^3_2 k^3_3}I(\mu)k_3^{3-4i\mu}\sum_{b=\pm} b \int^0_{-\infty} d\tau_1 \tau_1^{-1-2i\mu}(1-ik_{3}\tau_1 b)e^{ik_{3}\tau_1 b}\times\\ 
	\times & \sum_{a=\pm} a \int^0_{-\infty} d\tau_2 \tau_2^{-1-2i\mu}(1-ik_{12}\tau_2 a-k_1k_2\tau_2^2)e^{ik_{12}\tau_2 a}+(\mu\rightarrow -\mu) ~, \\ 
	\xrightarrow{k_1\simeq k_2\gg k_3}\simeq &
 	-\frac{1}{2} \lambda^2 \bar\chi   \frac{1}{  \pi^4  }  \frac{H^2}{64 k_1^3  k_1^3  }   \bigg[ I(\mu) \mu^{-2} (2-i\mu)  \Gamma^2 (2-2 i \mu )\sin^2(i\mu\pi) \bigg(\frac{k_3  }{2 k_1}\bigg)^{-2i\mu}  +\text{c.c.} \bigg] ~.
\end{align}
 
\paragraph{Three-point function with $\mathbf \lambda_2$:}
The three-point function \eqref{threepointcorrelationofchid} for $\lambda_2$ derivative coupling case
\begin{align} \nonumber 
	\langle \delta\chi_{\mathbf k_1} \delta\chi_{\mathbf k_2} \delta\chi_{\mathbf k_3} \rangle'_{(2)} =  &\frac{1}{2} (i \lambda_2)^2 \dot{\bar\chi} \sum_{a,b=\pm} ab \int \frac{d \tau_1}{(-H\tau_1)^3}   \frac{d \tau_2}{(-H\tau_2)^2} G^\prime_a (k_1;\tau_2) G^\prime_a (k_2;\tau_2) \times
 \\ \nonumber	 \times &
 G^\prime_b (k_3;\tau_1)\mathcal{I}(k_3;\tau_1,\tau_2)\\ \nonumber
	\xrightarrow{\tau_1,\tau_2\rightarrow 0}\simeq & \frac{\lambda_2^2 \dot{\bar\chi}}{2}H^{-5}  \frac{H^6}{8 k_1 k_2 k_3}\frac{H^4}{4\pi^4} \int^0_{-\infty}  d \tau_1 \int^0_{-\infty} d \tau_2\ \tau_1^{-3} \tau_2^{-2}  \sum_{a,b=\pm} ab \ \tau_2^2 \tau_1 \times\\ 
\times & e^{i a k_{12}\tau_2}e^{i b k_3\tau_1}I(\mu)(\tau_1\tau_2)^{3-2i\mu}k_3^{3-4i\mu}+(\mu\rightarrow -\mu) ~,
\end{align}
where $k_{12}=k_1+k_2$, 
\begin{align}\nonumber
	\langle \delta\chi_{\mathbf k_1} \delta\chi_{\mathbf k_2} \delta\chi_{\mathbf k_3} \rangle'_{(2)}\simeq & \frac{\lambda_2^2 \dot{\bar\chi}}{2^6 \pi^4}  \frac{H^5}{ k_1 k_2 k_3}{}I(\mu)k_3^{3-4i\mu}\sum_{b=\pm} b \int^0_{-\infty} d\tau_1 \tau_1^{1-2i\mu}e^{ik_{3}\tau_1 b}\times\\ 
	\times & \sum_{a=\pm} a \int^0_{-\infty} d\tau_2 \tau_2^{3-2i\mu}e^{ik_{12}\tau_2 a}+(\mu\rightarrow -\mu) ~, \\ 
	\xrightarrow{k_1\simeq k_2\gg k_3}\simeq & \frac{-\lambda_2^2 \dot{\bar\chi}H^5}{2^8  \pi^4  k_1^6}   \bigg[ I(\mu) \Gamma(2-2 i \mu )\Gamma(4-2 i \mu )\sin^2(i\mu\pi) \bigg(\frac{k_3  }{2 k_1}\bigg)^{-2i\mu}  +\text{c.c.} \bigg] ~.
\end{align}
  
\paragraph{Four-point function with $\mathbf \lambda$:}
The four-point function \eqref{fourpointcorrelationofchi} for $\lambda$ coupling case
\begin{align}\nonumber
	\langle \delta\chi_{\mathbf k_1} \delta\chi_{\mathbf k_2} \delta\chi_{\mathbf k_3} \delta\chi_{\mathbf k_4} \rangle'_{(2)} =& \frac{1}{2} (i \lambda)^2  \sum_{a,b=\pm} ab \int \frac{d \tau_1}{(-H\tau_1)^4}   \frac{d \tau_2}{(-H\tau_2)^4} G_a (k_1;\tau_2) G_a (k_2;\tau_2) \times
\\ \nonumber \times &
G_b (k_3;\tau_1) G_b (k_4;\tau_1) \mathcal{I}(k_I;\tau_1,\tau_2) ~,
\\ \nonumber
	\xrightarrow{\tau_1,\tau_2\rightarrow 0}\simeq \frac{- \lambda^2 }{2} H^{-8} \frac{H^8}{16 k^3_1 k^3_2 k^3_3 k^3_4} & \frac{H^4}{4\pi^4} \int^0_{-\infty}  d \tau_1 \int^0_{-\infty} d \tau_2\ \tau_1^{-4} \tau_2^{-4}  \sum_{a,b=\pm} ab(1-i b k_3\tau_1)(1-i b k_4\tau_1)\times\\ 
\times (1-i a k_1\tau_2)&(1-i a k_2\tau_2)e^{i a k_{12}\tau_2}e^{i b k_{34}\tau_1}I(\mu)(\tau_1\tau_2)^{3-2i\mu}k_I^{3-4i\mu}+(\mu\rightarrow -\mu) ~,
\end{align}
where $k_{12}=k_1+k_2$, $k_{34}=k_3+k_4$, and $\mathbf k_I=\mathbf k_1+\mathbf k_2=\mathbf k_3+\mathbf k_4$
\begin{align}\nonumber
	\langle \delta\chi_{\mathbf k_1} \delta\chi_{\mathbf k_2} \delta\chi_{\mathbf k_3} \delta\chi_{\mathbf k_4} \rangle'_{(2)}\simeq & \frac{- \lambda^2 }{2 \pi^4} \frac{ H^4}{64 k^3_1 k^3_2 k^3_3 k^3_4}I(\mu)k_I^{3-4i\mu}\sum_{b=\pm} b \int^0_{-\infty} d\tau_1 \tau_1^{-1-2i\mu}(1-ik_{34}\tau_1 b-k_3k_4\tau_1^2)\times\\ 
	\times e^{ik_{34}\tau_1 b} & \sum_{a=\pm} a \int^0_{-\infty} d\tau_2 \tau_2^{-1-2i\mu}(1-ik_{12}\tau_2 a-k_1k_2\tau_2^2)e^{ik_{12}\tau_2 a}+(\mu\rightarrow -\mu) ~, \\ 
	\xrightarrow[k_3\simeq k_4\gg k_I]{k_1\simeq k_2\gg k_I}\simeq \frac{-\lambda^2}{ 2^9 \pi^4 } \frac{H^4 k_I^3}{ k_1^6 k_3^6} & \bigg[ I(\mu) \mu^{-2} (2-i\mu)^2 \Gamma^2 (2-2 i \mu ) \sin^2(i\mu\pi) \bigg(\frac{k_I^2 }{4 k_1 k_3}\bigg)^{-2i\mu}  +{\rm c.c.} \bigg] ~.
\end{align}
 
\paragraph{Four-point function with $\mathbf \lambda_2$:}
The four-point function \eqref{fourpointcorrelationofchid} for $\lambda_2$ derivative coupling case
\begin{align}\nonumber
	\langle \delta\chi_{\mathbf k_1} \delta\chi_{\mathbf k_2} \delta\chi_{\mathbf k_3} \delta\chi_{\mathbf k_4} \rangle'_{(2)} =& \frac{1}{2} (i \lambda_2)^2  \sum_{a,b=\pm} ab \int \frac{d \tau_1}{(-H\tau_1)^2}   \frac{d \tau_2}{(-H\tau_2)^2} G^\prime_a (k_1;\tau_2) G^\prime_a (k_2;\tau_2) \times
\\ \nonumber \times &
G^\prime_b (k_3;\tau_1) G^\prime_b (k_4;\tau_1) \mathcal{I}(k_I;\tau_1,\tau_2) ~,
\\ \nonumber
	\xrightarrow{\tau_1,\tau_2\rightarrow 0}\simeq \frac{- \lambda_2^2 }{2} H^{-4} \frac{H^8}{16 k_1 k_2 k_3 k_4} & \frac{H^4}{4\pi^4} \int^0_{-\infty}  d \tau_1 \int^0_{-\infty} d \tau_2\ \tau_1^{-2} \tau_2^{-2}  \sum_{a,b=\pm} ab\ \tau_1^2\tau_2^2 \times\\ 
\times &e^{i a k_{12}\tau_2}e^{i b k_{34}\tau_1}I(\mu)(\tau_1\tau_2)^{3-2i\mu}k_I^{3-4i\mu}+(\mu\rightarrow -\mu) ~,
\end{align}
where $k_{12}=k_1+k_2$, $k_{34}=k_3+k_4$, and $\mathbf k_I=\mathbf k_1+\mathbf k_2=\mathbf k_3+\mathbf k_4$
\begin{align}\nonumber
	\langle \delta\chi_{\mathbf k_1} \delta\chi_{\mathbf k_2} \delta\chi_{\mathbf k_3} \delta\chi_{\mathbf k_4} \rangle'_{(2)}\simeq & \frac{- \lambda_2^2 }{2 \pi^4} \frac{ H^4}{64 k^3_1 k^3_2 k^3_3 k^3_4}I(\mu)k_I^{3-4i\mu}\sum_{b=\pm} b \int^0_{-\infty} d\tau_1 \tau_1^{3-2i\mu}(1-ik_{34}\tau_1 b-k_3k_4\tau_1^2)\times\\ 
	\times e^{ik_{34}\tau_1 b} & \sum_{a=\pm} a \int^0_{-\infty} d\tau_2 \tau_2^{3-2i\mu}(1-ik_{12}\tau_2 a-k_1k_2\tau_2^2)e^{ik_{12}\tau_2 a}+(\mu\rightarrow -\mu) ~, \\ 
	\xrightarrow[k_3\simeq k_4\gg k_I]{k_1\simeq k_2\gg k_I}\simeq \frac{\lambda_2^2}{ 2^{13} \pi^4 } \frac{H^4 k_I^3}{ k_1^6 k_3^6} & \bigg[ I(\mu) \Gamma^2 (4-2 i \mu ) \sin^2(i\mu\pi) \bigg(\frac{k_I^2 }{4 k_1 k_3}\bigg)^{-2i\mu}  +{\rm c.c.} \bigg] ~.
\end{align}
\color{black}
\bibliography{superheavyDM}{}
\bibliographystyle{utphys}

\end{document}